\documentclass[11pt]{article}
\usepackage{amsmath}
\usepackage{amsfonts}
\usepackage{graphicx}
\usepackage{hyperref}
\usepackage{url}
\usepackage{xcolor}
\usepackage{microtype}
\usepackage{booktabs}
\usepackage{fancyhdr}
\usepackage{lipsum}
\usepackage{siunitx}
\usepackage[T1]{fontenc}
\usepackage[a4paper, left=2.1cm, right=2.1cm]{geometry}

\newcommand{\E}{\operatorname{e}}

\newcommand{\D}{\mathop{}\!\mathrm{d}}
\newcommand{\Dt}[1][t]{\D#1}

\title{Reducing Cartel Violence: The Mexican Dilemma\\ Between Social and Security Spending}

\author{
  Rafael Prieto-Curiel\thanks{Complexity Science Hub, Metternichgasse 8, Vienna, Austria. \texttt{prieto-curiel@csh.ac.at}} \and
  Dieter Grass\thanks{International Institute for Applied Systems Analysis (IIASA), Laxenburg, Austria} \and
  Stefan Wrzaczek\footnotemark[2] \and
  Gian Maria Campedelli\thanks{Fondazione Bruno Kessler, via Sommarive 17, Povo, Trento, Italy} \and
  Gernot Tragler\thanks{TU Wien, Research Unit of Variational Analysis, Dynamics and Operations Research (VADOR), Vienna, Austria} \and
  Gustav Feichtinger\footnotemark[4]
}

\date{}

\begin{document}
\maketitle

\section{Abstract}

Organised crime in Mexico threatens societal stability and public safety, driving pervasive violence and economic disruption. Despite security investments and social programs designed in part to reduce involvement in crime, cartel power and violence continue to persist. This study evaluates existing policies and introduces a novel framework using optimal control theory to analyse cartel dynamics. Specifically, by modelling resource allocation between security measures and social programs, we identify optimal strategies to mitigate the impacts of cartels. Findings reveal that Mexico's largest cartel imposes an annual economic burden exceeding \text{US\$ } 19 billion, 2.5 times the government's investment in science and technology. We further demonstrate that current budget allocations between social and security programs are nearly optimal yet insufficient to reduce cartel violence significantly. In light of these findings, we demonstrate that achieving meaningful harm reduction would require a significantly larger budget and would take over a decade, even with increased funding.

\section{Introduction}

{
Approximately one in three intentional homicides worldwide occurs in Latin America, despite the region representing only 8\% of the global population \cite{UNODC}. Beyond the human toll, organised crime generates pervasive fear and displacement, affecting millions across the region \cite{LAPOP, MigrationConflictColombia}. Mexico, in particular, has been severely impacted, with homicides increasing by over 300\% between 2007 and 2021, making it one of the most violent countries globally that is not at war \cite{UNODC}. The economic impact of violence is staggering, estimated at almost 20\% of Mexico's GDP \cite{jaitman2017costs}. Violence remains the country's top concern, with nearly half of the population blaming cartels \cite{Latinobarometro2023}. Yet, the burden of cartels is not exclusive to Latin America. Mexican cartels have also established transnational networks, fueling a global public health crisis by supplying illicit drugs. In the US, their largest market, drug overdoses claim approximately 100,000 lives annually, a figure that has doubled in a decade \cite{friedman2022charting, klobucista2023fentanyl}.
}

{
Given the far-reaching societal costs posed by organised crime groups, institutions in Mexico attempt to curb their. The economic investment made by Mexico in terms of national security programs each year amounts to \text{US\$ }9.88 billion \cite{PEF2023}, seeking to increase the strength of the army, marines, police, and other bodies aimed at targeting crime (see the Supplementary Material SM - A). However, existing strategies have failed to contain violence, reduce the power of cartels, block the flow of weapons, or prevent access to illicit substances \cite{UNODCDrugs2022}. Targeting criminal leaders \cite{velasco2023unintended, estevez2024impact}, antigang crackdowns \cite{lessing2016inside, wolf2017mano}, gun and drug seizures \cite{shirk2015understanding}, mass incarcerations \cite{lessing2017making} and the destruction of illegal structures have been central elements of the security strategy of both countries for decades \cite{TrejoWhyDidDrug2018, TrejoHighProfileCriminalViolence2021}. 
}

{
In parallel, Mexico has also deployed a broad set of social programs, including community centres \cite{vilalta2025beyond}, sports programs \cite{jugl2021sports}, and cash transfers for all adults above a certain age, single mothers, students, farmers and more \cite{martinez2023politica}. Although the core objective of social policies is to improve social welfare (reduce inequality, increase school enrollment rates, and reduce poverty), they are often portrayed as part of the government's strategy to deter young men from joining criminal gangs \cite{economist2025mexico}. Social programs are an instrument to indirectly counter involvement in criminal activities because, with a higher monetary value of a social program, there exist fewer relative gains and, thus, motivations to join cartels \cite{ehrlich1973participation, freeman1999economics}. Each year, \text{US\$ }8.84 billion is invested in these social programs \cite{PEF2023}, reaching millions of beneficiaries (SM - A). 
}

{
Despite the vast economic costs of security and social programs, violence is rising in Mexico, while drug overdose deaths in the US have been a national emergency \cite{ScienceCommentaryCaulkins}. In this evolving landscape, and despite the impact cartels have within and well beyond Mexico, we lack information on how to allocate expenditures efficiently. Are social programs effective in preventing cartel recruitment and, therefore, future violence? If so, should portions of the security budget be reallocated to social programs? Alternatively, if security programs effectively reduce violence, should parts of the social budget be allocated to security programs? Should more resources be spent on both security and social programs? 
}

{
Here, we attempt to address these questions by analysing cartel dynamics in Mexico and forecasting cartel violence across a set of policy scenarios within an optimal control framework. Optimal control theory has been widely applied to investigate phenomena such as attitudes towards crime reporting \cite{Short2012ExternalConversions}, terrorist behaviours \cite{caulkins2009optimal}, and illicit drug demand \cite{tragleretal2001}. The intuition behind this approach is that the outcome of policies (or controls) and their associated costs can be expressed in the form of an optimisation problem. Optimal control theory enables the examination of dynamic policy solutions that evolve in response to the system's internal dynamics as its state changes. Here, the optimisation problems are formalised such that the budget can be allocated to social programs or used to counter different cartels through security expenditures. Considering the limited budget of Latin American countries, detecting how much to spend and how to allocate economic resources is a critical issue \cite{izquierdo2018better}. By assigning a monetary cost to future casualties and the harm of cartels and by considering the budget for social programs and security, those outcomes can be regarded as alongside and balanced against the budgetary costs for social programs and security, and so shed light on how resources should be best allocated to counter the influence and violence generated by cartels effectively. 
}

\subsection{Modelling cartel dynamics and countermeasures}

{
The total number of cartel-affiliated individuals in Mexico has recently been estimated at 175,000 \cite{PrietoCampedelliHope2023}. Approximately, they recruit 19,000 new members and experience about 6,000 incapacitations and 7,000 losses due to inter-cartel killings, disappearances, or retirements each year, resulting in a net annual increase of around 6,000 members \cite{PrietoCampedelliHope2023}. Building on those estimates, we determine the optimal strategy through a cost-benefit analysis, balancing the expenditure on social programs against security. While cartel size and the number of homicides are correlated, they conceptually refer to distinct objectives from the state's standpoint. Cartel conflicts often involve terrorist tactics and advanced weaponry, driven by retaliatory cycles that escalate until one group is defeated, displaced, or fragmented into smaller cells \cite{calderon2021organized, jones2018strategic}. Many cartel-related homicides occur when cartels fight against each other. Therefore, this first indicator is modelled as being proportional to the product of the two cartels' sizes. We refer to it as \textit{homicides of cartel members by other cartels}, but it also includes the kidnapping, torture, and killing of members of the opposing cartel (not of the general public). If the objective is to reduce the number of homicides, having a single, but large, cartel may be theoretically more desirable than having two conflicting cartels. The second indicator is broader, encompassing a myriad of other harms caused by cartels and suffered by the public more generally. It is assumed to be proportional to the size of the cartel and includes crimes such as the kidnapping and killing of journalists and other citizens, corruption of government officials, subversion of elections, and various economic offences. These economic activities include drug distribution, migrant smuggling, petroleum and ore theft, and the imposition of ``protection fees'' and other forms of extortion on legitimate businesses—famously including even avocado production and export. 
}

{
\paragraph{Modelling cartel size variation.} Due to their secretive nature, many dimensions of cartels remain unknown \cite{contreras2024computational}, but \cite{PrietoCampedelliHope2023} recently proposed a model that captures why and to what extent the size of cartels varies over time. According to this model, cartels grow because they recruit new members, but shrink due to incapacitation by the state, conflict with other cartels, and internal instability. Specifically, the number of members of cartel $i$ at time $t$ measured in weeks (where the explicit time dependence is suppressed in several parts of the text), expressed as $C_i(t)$, produces a system of coupled differential equations, one for each cartel: 
\begin{eqnarray}\label{eq:MasterEquation}
\dot{C}_i (t) & = & \underbrace{g(v(t), C_i (t))}_\text{recruitment} - \underbrace{f(u_i (t), C_i (t))}_\text{incapacitation} - \underbrace{\theta C_i (t) C_j (t)}_\text{conflict} - \underbrace{\omega C_i^2 (t)}_\text{saturation}. 
\end{eqnarray}
The term $\theta C_i C_j$ captures the lethality $\theta$ of the conflict between cartels $i$ and $j$. The term $\omega C_i^2$ captures internal instability (group dropouts, disputes, retirement, and diminishing returns for large groups). We model the impact of social programs with $g(v, C_i) = \rho b \E^{-\sigma v} C_i$, capturing how cartel recruitment is reduced when a person receives $\text{US\$ } v$, depending on the \textit{efficiency} of social programs $\sigma$, the recruitment rate $\rho$ and a baseline rate $b$. The impact of security programs is modelled with $f(u_i, C_i)= \eta \left(u_i C_i \right)^\pi$, which maps how the budget $\text{US\$ }u_i$ used for fighting cartel $i$ via security expenditures by the state, depending on the \textit{efficiency} of security programs, $\pi<1$, and a baseline $\eta$. We theorise that with a higher security budget, $u_i$, cartel $i$ will have more incapacitations. However, based on the principle of optimal foraging \cite{vandeviver2023foraging}, incapacitations become progressively more challenging (SM - C). Therefore, with higher values of $u_i$, the function $f$ increases, but at a diminishing rate.
}

{
\paragraph{Estimating the scale and effectiveness of spending on social welfare.} Cartels offer premium salaries and other perks to attract new members \cite{SVyP2024TikTok}, often through social media posts \cite{bledsoe2025socialmedia}. We model both forced and voluntary recruitment, accounting for how legal income and social programs may deter participation (details in the Methods). When a person is exposed to voluntary recruitment, they compare the \textit{cartel offer}, which includes higher income but also the risk of being incapacitated or killed by rival cartels, with the \textit{civil offer}, the combined income from legal employment and social programs. If exposed to forced recruitment, the person joins the cartel without choice. If exposed to voluntary recruitment, they join with a probability that decreases as the civil offer approaches the cartel offer. Therefore, as the value of a social program, $v$, increases, cartel recruitment becomes less frequent, as captured by the decrease in the recruitment function, $g$. In 2023, nearly 50\% of the families of the country reported receiving some aid directly from the state \cite{Latinobarometro2023}. However, social programs directed towards vulnerable individuals to prevent cartel recruitment reach approximately $P = 10.7$ million \cite{PEF2023}. The budget used for those social programs is $\phi =P v = \text{US\$ }8.84$ billion each year. A social program provides, on average, a yearly cash transfer of $v_0 = \text{US\$ }832 $ per person. 
}

{
\paragraph{Estimating the scale and effectiveness of government spending on security.} We conduct a cost-benefit analysis of social and security programs, taking into account the burden created by cartels. We assume that a large share of the burden of violence is related to cartels and that the same share of the security budget is used to fight against them. We take this share to be 80\%, so of the burden of violence, 80\% is related to cartels (and the remaining 20\% is associated with other types of crime) and of the security budget, 80\% is used to fight cartel members (and the remaining 20\% is used to fight different kinds of crime). We will vary this assumed share as part of our sensitivity analysis. In Mexico, the security budget amounts to $\text{US\$ }9.88$ billion yearly \cite{PEF2023}. We focus on the conflict between the two largest cartels ($C_1$ - Cartel Jalisco Nueva Generación and $C_2$ - Cartel Sinaloa), identified as the most relevant security challenges in the country \cite{paoli2024sinaloa, pereda2024illegal} and assume that the conflict between other cartels follows a similar dynamic in terms of outcome. Based on the cartel size estimate provided by \cite{PrietoCampedelliHope2023}, we assume that the budget against organised crime is homogeneously distributed across cartels depending on their size. Thus, $\text{US\$ }1.11$ billion each year is used to fight directly $C_1$ and the same against $C_2$, assuming that both cartels have a similar size today (SM - B).
}

{
\paragraph{Budget allocation and objective function.} We consider the budget spent on security programs and the budget for social programs as the policy tools (the \textit{controls} in mathematical terms) that the decision-maker has available to reduce the burden created by cartels. We design our approach such that enforcement can target either cartel, but social spending is sent to all eligible people, and so has the same proportionate effect on recruitment by both cartels. Hence, there are two distinct security spending variables, one for each cartel, but only one variable describing spending on social programs. We have a budget restriction given by how much was spent on all three policy tools combined:
\begin{equation} \label{BudgetRestrictions}
\underbrace{u_1 (t) + u_2 (t)}_\text{security budget} + \underbrace{P v (t)}_\text{social budget} \leq \text{US\$ }11.1 \text{ billion each year.}
\end{equation}
}

{
Combining the harm that cartels create with the cost of control spending, and planning over a time horizon (with a standard approach to discounting future outcomes), we express the government's objective as 
\begin{eqnarray} \label{ObjectiveEq}
Q&:=& \int_0^\infty \E^{-rt} \left( \underbrace{2 s \theta C_1(t) C_2(t)}_{\text{homicides}} + \underbrace{ h ( C_1(t) + C_2(t))}_{\text{cartel harm}}+ \underbrace{ u_1(t) + u_2(t)}_{\text{security}} + \underbrace{P v(t)}_{\text{social}} \right) \,\Dt,
\end{eqnarray}
where $s$ is the cost assigned to a cartel member being murdered by another cartel, whilst $h$ encapsulates the cost of all other cartel-related harms (SM - A). In line with the frequent short-term horizon of political objectives, we consider a weekly discount rate of $r = 0.001$ per week (roughly 5\% per year), so the future costs are less relevant than the present for the policymaker \cite{jacobs2011governing}. The expression \ref{ObjectiveEq} takes into account the costs of the controls (security and social programs) as well as their impact (reducing homicides and cartel harm). Hence, it identifies which policy actions (reducing or increasing any of the controls) are more effective in achieving the overarching goal of combating cartels. 
}

\section{Results} 

{
First, we analyse the scale of harm. It is challenging to quantify the burden of violence in Mexico, but it has been estimated at \text{US\$ }230 billion in 2022 \cite{index2023mexico}. That figure includes indirect costs, such as a reduction in tourism, a loss in productivity, displacement of people from their homes, a lack of investment, and deterioration of public life, among many others \cite{aburto2023global, cervantes2023estimating, BIDCostosViolencia}. Of the estimated costs, 45\% was attributed to the 31,000 intentional homicides in the country \cite{index2023mexico}, so we assign to each average value of \text{US\$ }3.3 million for each life lost. Further, it was estimated that the fight between the top two cartels has resulted in the murder of about 780 of their members each year, which would multiply out to a cost of \text{US\$ }2.6 billion \cite{PrietoCampedelliHope2023}. Similar to the budget assignment to cartels, we assume that 80\% of this burden is attributable to cartels, so the yearly impact of each member is $h = \text{US\$ }0.58$ million.
}

{
With roughly 25,000 members, we estimate that the biggest cartel in Mexico creates a yearly burden of $\text{US\$ }19.1$ billion for the country. Of all costs produced by $C_1$, 76\% are related to their harm, 18\% are due to the homicides they perpetrate against other cartels and only 6\% to the security budget used to fight against them (Figure \ref{Scheme}). To contextualise these figures, the burden of $C_1$ is roughly 2.5 times what the government invests in Science and Technology \cite{PEF2023}. 

\begin{figure}[!htbp]
\centering
\includegraphics[width=0.75\textwidth]{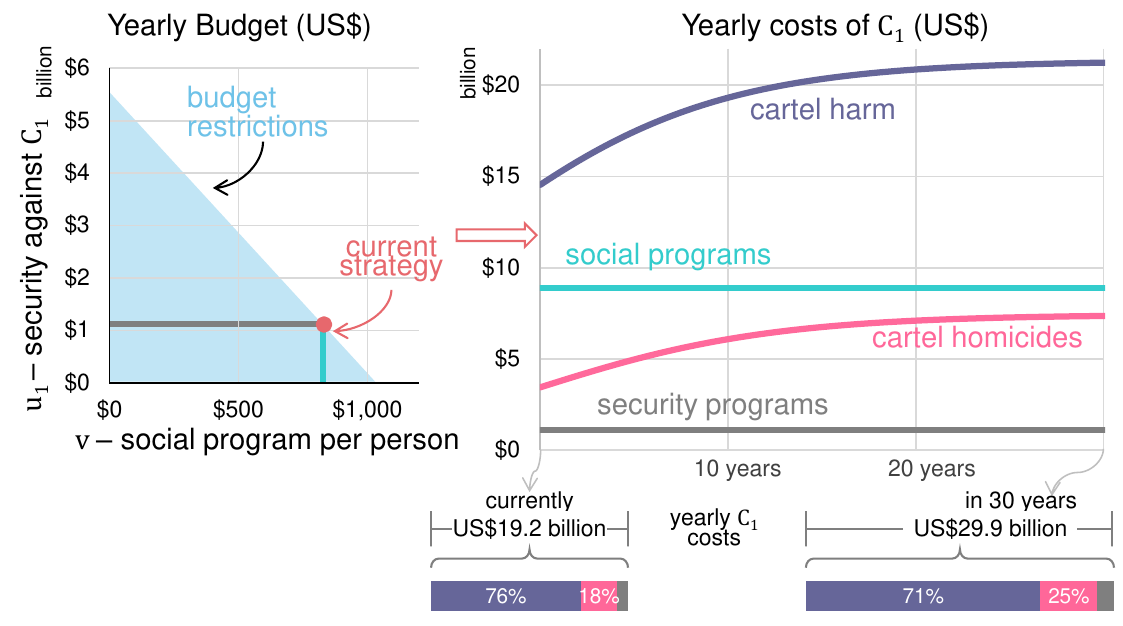}
\caption{Left - Value of a social program $v$ (horizontal) and security budget against $C_1$, $u_1$ (vertical) which combine into a budget restriction. Right - The estimated current cost of $C_1$ is $\text{US\$ }19.1$ billion, 76\% due to their harm, 18\% related to their homicides and 6\% related to the security budget used against them. The modelled costs of $C_1$ in 30 years are 56\% higher due to an increase in cartel harm and homicides. }\label{Scheme}
\end{figure}
}

{
\paragraph{Forecasting cartel homicides and harm.} To establish a baseline, we first visualise what our model indicates would happen if spending were maintained at current levels. With the current budget allocated to security and social programs, we project that the top two cartels will continue to grow (Figure \ref{ResultsFigure30years}). Cartel homicides and their harm would increase, reaching within a decade 15\% more homicides than those observed today. The 30-year cost of only the top two cartels in terms of the homicides they produce against each other, their harm, and the budget spent to stop them via security and social programs is thus estimated to be over $\text{US\$ }804$ billion. Although many exogenous shocks and other factors not considered by the model may alter the future, we use these estimates as a benchmark against other budget strategies (SM - D). 

\begin{figure}[!htbp]
\centering
\includegraphics[width=.85\textwidth]{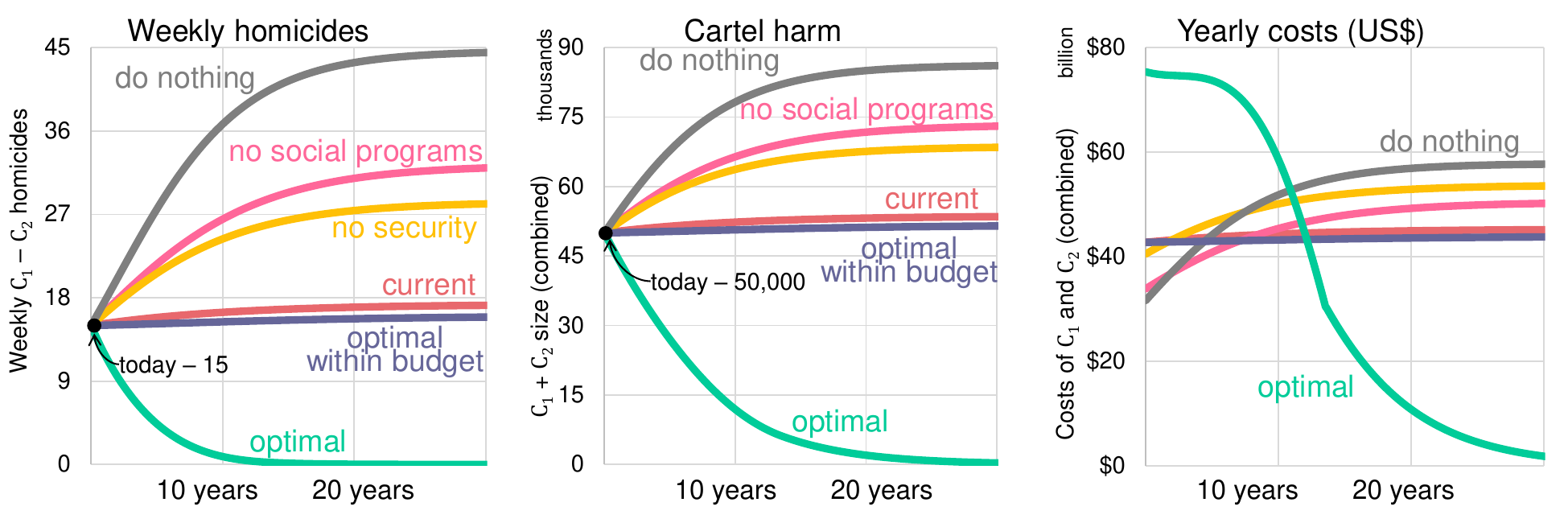}
\caption{Cartel homicides between $C_1$ and $C_2$ in the next three decades, depending on the budget used for social programs and the budget used for security programs (left). Cartel harm quantified by the combined size of $C_1$ and $C_2$ in the next three decades (centre). Total costs are considered, including social programs, security programs, the cost of homicides, and the cost of cartel harm (right). }\label{ResultsFigure30years}
\end{figure}
}

{
\paragraph{Varying choices of budget assignment.} There is frequent debate over the impact of social and security programs. We thus analyse what the model projects when the total existing budget was devoted (1) just to social programs, (2) just to enforcement, or (3) they were reallocated to other priorities having nothing to do with cartels. We estimate that if the security budget that is used to fight $C_1$ and $C_2$ were re-assigned to social programs, a beneficiary would receive up to \text{US\$ }1,040 each year, and that would reduce current cartel recruitment by about 6\%, but dropping incapacitations to zero. Similarly, if the budget for social programs were used for security and distributed among the top two cartels, the yearly budget for fighting against $C_1$ would be $\text{US\$ }5.55$ billion (increasing incapacitations by 120\%). However, by removing social programs, cartel recruitment would increase by 26\%. This illustrates the tradeoff embodied in the model between spending more on social programs and spending more on security. All of those counterfactuals are notably worse than the original projection with the current policy with respect to all three outcomes (cartel-on-cartel homicides, cartel harm, and total social welfare). The only exception is that giving up and doing nothing saves money for the first three years, but after that, the cartels have grown enough to leave society worse off than with the current policy (Figure \ref{ResultsFigure30years}). The foregoing analysis suggests that the current division of cartel-control spending is worthwhile, and it outperforms either a purely security or a strictly social programming approach (SM - D). The benefits of constraining cartel size are at least partially greater than the costs. Furthermore, the current mix of program spending fares favourably relative to either allocating all funds to enforcement or all to social programs.
}

{
We compute the optimal dynamic budget allocation that reduces the long-term costs of violence and the use of security or social programs (SM - F). According to our model, that strategy would invest significant amounts initially in security programs  (with $u_i(0) = \text{US\$ }2.4$ billion each year) and a smaller amount on social programs (with $v(0) = \text{US\$ }582$ per person) and they would slowly move to a smaller budget on security (with $u_i(10\text{ years}) = \text{US\$ }1.8$ billion) and more on social programs (with $v(10\text{ years}) = \text{US\$ }697$). However, even in the optimal assignment, both cartels would keep growing, expanding their homicides and harm. The 30-year cost of the optimal assignment of the budget is only 0.8\% smaller than the current costs, without any significant drop in cartel violence. Given the current total amount of economic resources invested in social and security spending, there is no significantly better allocation strategy to reduce homicides or curb cartel harm and in a few years, the model projects that Mexico will have more homicides and cartel harm than today (SM - I). Cartels are dynamic organisations, capable of adjusting their strategy in response to changing conditions. For example, if social programs increase, cartels can compensate by increasing the premium they offer to maintain their recruitment. This, however, significantly increases its costs, reducing its revenue and optimal size (details in the Methods). Even if cartels adjust their strategy, they are incentivised to downsize their operations, creating less harm and fewer homicides (SM - G). 
}

{
\paragraph{From budget allocation to budget increase.} We investigate how an unconstrained budget (relaxing from the inequality in Eq. \ref{BudgetRestrictions}) would affect cartel violence and harm, seeking to understand whether Mexico should increase its overall amount of resources used to counter the two leading criminal organisations of the country effectively. If the government invested an unconstrained budget in reducing its violence, the optimal strategy would increase social programs substantially, granting each person nearly 3.1 times more money (approximately $\text{US\$ }2,600$ per year). Social programs and the salary from legal activities could be more attractive than the premium and the risks offered by the cartel (details in the Methods). With higher social programs, we model a recruitment rate 40\% smaller than its current levels. Under the optimal strategy, social programs would expand significantly beyond today's levels and remain elevated for more than a decade. Then, after the cartels have been reduced in size, they would be shrunk after approximately 14 years. However, with an unconstrained budget, the most significant change is the increase in security programs. Investment in these programs would initially rise to 7.2 times the current level and continue to increase. Within a decade, the security budget is expected to peak at approximately 16 times the current investment, reflecting a significant escalation in the resources required for security (Figure \ref{OptimalBudgetStrategies}). We estimate that the 30-year cost of the optimal plan would save 11\% of the burden of violence. Although certain investments, particularly in security, can lead to short-term spikes in violence \cite{dell2015trafficking}, we have modelled and quantified these shocks, and they are relatively minor over 30 years (see Methods for details). Additionally, some programs may cause cartel fragmentation, where a cartel splits into independent and often rival factions \cite{calderon2015beheading}. While this fragmentation typically reduces overall cartel-related harm, it frequently leads to an increase in cartel-related homicides, which nearly offsets the benefit (see Methods). Initially, the budget required for social and security programs is four times what is invested today and would peak at 5.3 times more in seven years. In this scenario, the initial costs are significantly higher, but the decrease in cartel harm and the reduction in homicides ultimately pay off. Results show that with an unconstrained budget, it becomes possible to tip the system from a high to a low-violence equilibrium.

\begin{figure}[!htbp]
\centering
\includegraphics[width=0.5\textwidth]{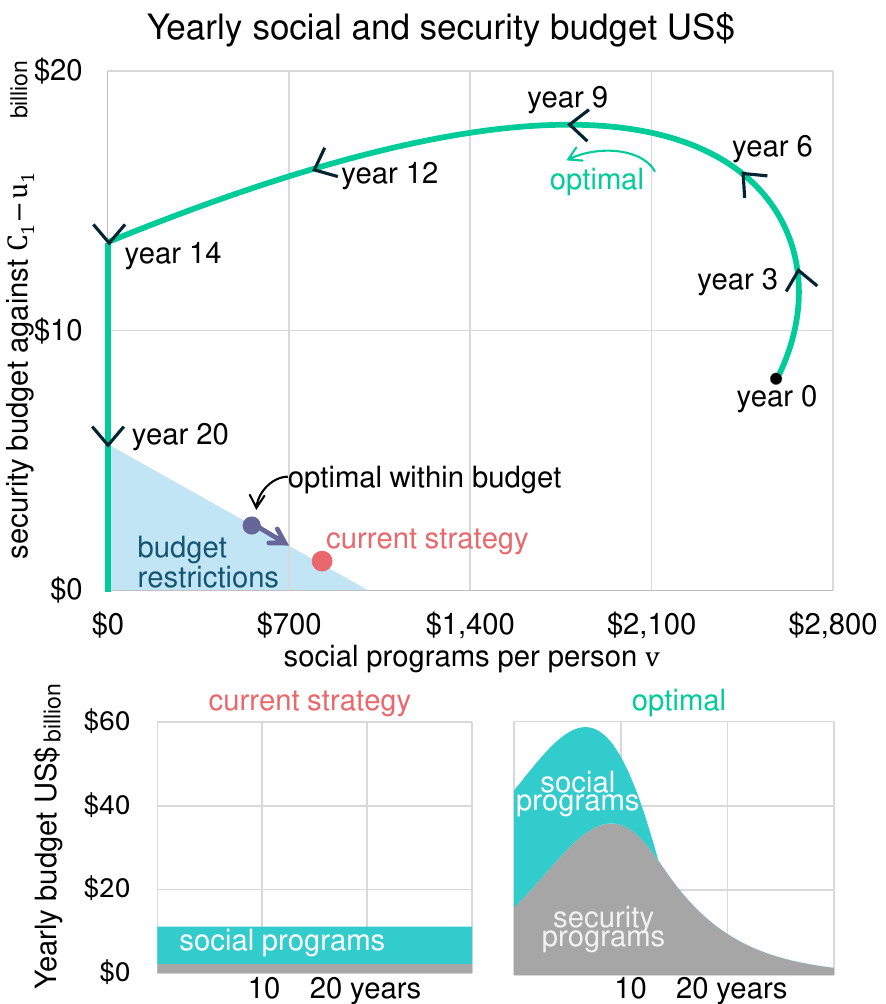}
\caption{Summary of budget strategies representing the controls on each axis. Budget strategies using social programs ($v$ in the horizontal axis) and security programs ($u_i$ in the vertical axis). Considering that $u_1 = u_2$, the budget restriction $u_1 + u_2 + Pv \leq \text{US\$ } 11.1$ billion is represented by the triangle. }\label{OptimalBudgetStrategies}
\end{figure}
}

{
With the current budget allocated to security and social programs, the number of homicides and overall cartel harm will increase (SM - E). However, with an unconstrained budget and an optimal strategy, reductions would eventually occur. It takes three years to cut the homicide rate in half and more than eight years to achieve a 90\% reduction in cartel-related homicides. Similarly, cartel harm would be reduced by half in nearly six years, while a 90\% reduction would require around 14 years.
}

{
According to the national victimisation survey, each year, over 30 million crimes are committed in Mexico \cite{ENVIPE2022}, yet only 120,000 individuals are incapacitated annually, primarily for minor offences \cite{ENPOL}. Mexico's impunity rate remains critically high, with intentional homicide impunity at 96\% in 2022 \cite{MexEvaluaImpunidad}. Security programs in Mexico could become more efficient at incarcerating criminals by prioritising the use of advanced police intelligence and focusing efforts on apprehending homicide perpetrators. Yet, one of the key results is that the outlook for the next decades does not rely heavily on how efficiently security resources are utilised (Figure \ref{StrategySummary}). If somehow security resources were more efficient (for example, aiming a specific target and managing to incarcerate more members using the same resources), then, with the current budget, we still expect an increase in cartel homicides and harm in the upcoming years (SM - H). Additionally, the optimal strategy would not differ much, and it would still take years to reduce violence.

\begin{figure}[!htbp]
\centering
\includegraphics[width=0.75\textwidth]{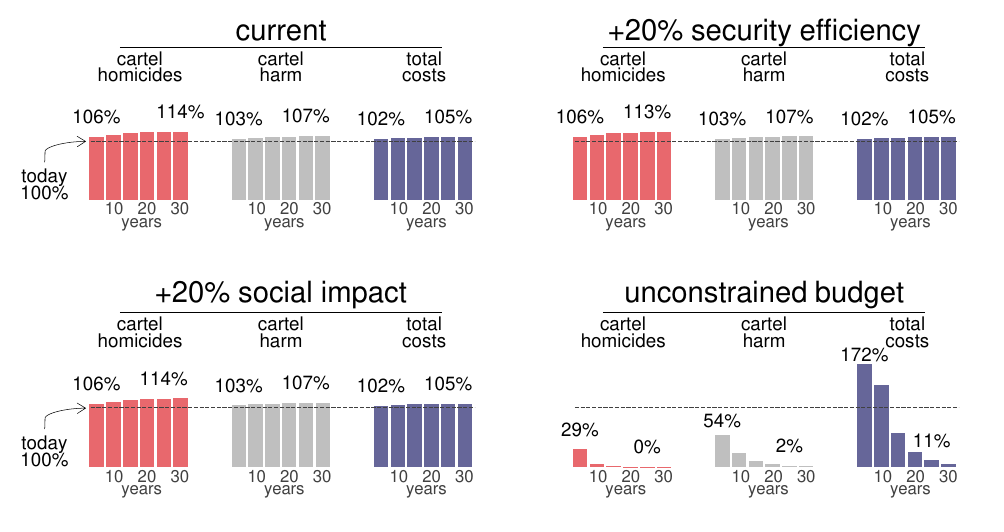}
\caption{Changes in the number of cartel homicides, combined cartel size and total costs, taking the first year as a baseline. }\label{StrategySummary}
\end{figure}
}

{
The cartel participation model shows that with higher income from legal activities (which include salaries and social programs), many people will still be exposed to recruitment but they will decide not to join a cartel (SM - D). Similarly, targeted social programs addressing vulnerable groups (e.g., areas with a significant cartel presence) could reduce cartel recruitment by providing sustainable alternatives and fostering long-term resilience in vulnerable communities. However, a similar result is observed in terms of the efficiency of social programs. If they were slightly more effective in preventing cartel recruitment, the country would still experience an increasing level of violence (Figure \ref{StrategySummary}). Partly, this is because the number of beneficiaries is too large, so the social programs are too costly, and still they provide only a fraction of what the cartel offers. Results show that social programs, with their current structure and number of beneficiaries, are not very effective and improving the efficiency only slightly (taken here to be $+20\%$) has almost no effect. Thus, enhancing security resources or strategically targeting social programs only has a limited impact if they continue to be constrained within the current budget (SM - G). 
}

{
Although the objective function quantified cartel harm and the homicides between cartels in monetary terms, the impact of the different strategies is also tangible in the number of lives lost. The model estimates that within the next ten years, the top two cartels will have inflicted against each other approximately 11,000 homicides. This number would increase by nearly 80\% if the government stops social and security programs, would remain roughly equal with more efficient security or social programs, but would drop by 74\% in the optimal scenario. 
}

{
Although the model relies on a few parameters, testing to ensure the robustness of its outcomes suggests that three key findings are robust to changes in its values: (1) the current budget for social and security programs is insufficient to shrink cartel size or violence, (2) at current budget levels, these programs alone cannot appreciably reduce violence in Mexico, but they significantly mitigate cartel impacts compared to having no programs, and (3) very substantial additional investment in security and social spending is necessary to tip the situation down to a different equilibrium with much smaller cartels. Furthermore, even with unconstrained resources, the optimal strategy would require years to reduce cartel harm and homicides and decades to eliminate the two central cartels effectively.
}

\section{Discussion}

{
Violence is having a devastating impact in Latin America and, in particular, in Mexico. Currently, 15 of the 50 most violent cities in the world are in Mexico (with Tijuana and Acapulco leading the list), with 43 of them being in Latin America and the Caribbean. State forces have invested financial and human resources for decades to reduce the burden of organised crime. Criminal incapacitations, drug seizures and dismantling of illegal structures have been part of the country's security strategy for decades \cite{atuesta2017meet, aguayo2017desamparo}. Yet, efforts have failed to contain violence, reduce the power of cartels, block the flow of weapons and have not prevented access to illicit substances in the US \cite{payan2013war, trejo2020votes}.
}

{
In this work, we introduced an optimisation model to estimate and study the dramatic economic burden of cartels in Mexico. We focused on the two central cartels, assuming that other large criminal organisations create a scaled burden to society. To enable comparison between the burden of cartels and the investments made in security and social programmes, we monetise indicators of violence, an approach that inevitably raises ethical concerns. Most of these costs are related to the harm of cartels; a smaller part is due to the homicides they create, and an even smaller share is associated with the security budget that is invested in fighting organised crime. Cartels impose a social burden that is measured in billions, but the country seeks to control them with a budget that corresponds to 6\% of that burden. This constraint is so severe that our model shows that even investing such a fixed amount of resources today in social and security programs will not reduce violence, but the trend in homicides is predicted to grow, along with broader cartel harms. 

}

{
The results presented here are the product of a model, and so they must be interpreted as a simplification and abstraction of a much more complex reality. We capture the harm of cartels only based on their size, but there are many dimensions. It does not account for political feasibility, administrative capacity, delays in policy implementation, the challenge of coordinating federal and state forces, or the limitations of the judicial and penal systems, despite their critical role in deterring crime. Distinguishing between investments in security forces and those in the justice system is essential, particularly given the widespread inefficiencies that characterise the current landscape. Also, regional disparities further complicate the picture. Mexico has significant regional disparities, between states like Yucatán, having homicide rates similar to Europe, and states like Colima that would rank among the most violent globally if they were their own countries. Conditions at the local level are highly heterogeneous, and so are the policies required. Security spending varies significantly across states and municipalities, shaped by local political coalitions and institutional weaknesses. These differences affect not just the amount spent but also how effectively resources are deployed. Local and regional governments often lack the oversight, capacity, or political will to implement evidence-based interventions. The model also treats cartels as homogeneous entities differentiated only by size. In reality, there is considerable heterogeneity across criminal organisations in terms of structure, territorial control, economic strategies, and political alliances. These differences likely influence not only the scale and nature of violence but also the efficacy of policy interventions. Thus, results need to be considered with caution. 
}

{
In spite of the limitations, the results within our modelling framework are robust, as they hold even after sensitivity tests are performed to assess the role that assumptions have on the final estimates, thus providing a first comprehensive empirical analysis that could shed light on the medium- and long-term consequences of Mexican policies and cartel violence. Here, we provide a cost-benefit analysis of social and security programs, where we quantify the output of each dollar spent as a control (either part of social programs or the security budget). To that expression, we add the burden of cartels through their harm and their homicides against other cartel members. That permits us to consider the impact of cartels as an optimisation problem in the framework of optimal control theory. Within the existing budget restrictions, the division between security and social spending has only minor impacts in terms of countering cartels. We estimate that dropping any of the programs results in even higher social costs. Without social programs, for example, cartels could recruit more members than they do today, becoming an even larger social threat. Similarly, without security programs, cartels would reach a much larger size, with a higher number of homicides and more harm. Thus, investing in security and social programs is essential to avoid an even larger spread of cartel power and prevalence in the country. Actually, within the framework of our model and the budget restrictions, the current assignment between social and security programs is not far from optimal. However, according to the model, Mexico's investments against cartels are too small. The country spends too little on security programs, particularly considering the size of its cartels and the burden they pose. We estimate that for every dollar that the government spends to fight the biggest cartel, the burden that it has is $\text{US\$ }15.6$, out of which $\text{US\$ }3$ are due to the homicides against other cartels and $\text{US\$ }12.6$ due to its broader harm. 
}

{
Focusing on the model's forecasts for violence levels over the next few years, results indicate that reducing cartel violence or reducing cartel harm is not feasible within the given budget constraints. But even with an unconstrained budget, cartels will have an impact on Mexico for years. A significantly larger budget is required for a faster reduction in cartel harm and homicides. For example, even under extreme values (far beyond our model's calibrated range), achieving a 95\% reduction in cartel-related homicides within a single year would necessitate an investment exceeding $\text{US\$ }2$ trillion, equivalent to 1.2 times Mexico's entire annual GDP. This would require increasing the current security budget by nearly a factor of 1,000, a scale of spending that is not only economically unfeasible but politically and logistically inconceivable.
}

{
Compared to most countries, Mexico is characterised by how little it spends on security. Despite suffering extremely high levels of violence, the country spent only 0.63\% of its GDP on the justice system and domestic security in 2021, roughly one-fourth of the budget spent by other countries in Latin America \cite{index2023mexico, OECDSpendingSecurity}. An interesting regional comparison, despite the structural, social, and political differences between the two, is Colombia. During the 1980s and 1990s, the country faced severe threats from organised crime, leftist guerrilla movements, and far-right paramilitary groups, all of which amassed thousands of members \cite{DeshazoBackBrinkEvaluating2007}. Moreover, the country's security institutions were severely under-resourced in terms of both manpower and funding. After peace negotiations with the FARC failed, President Pastrana launched a major military offensive against the group in 2000, backed by financial and logistical support from the United States. The resulting initiative, known as \textit{Plan Colombia}, continued until 2015 and led to substantial investments in equipment, training, and security infrastructure. The plan is widely considered a partial success, with its most notable achievement being a dramatic reduction in violence across the country \cite{Felbab-Brownviolentdrugmarket2009}. Today, Colombia invests roughly four times the share of its GDP in domestic security compared to earlier periods, and has reduced its violence by approximately 66\% from 1990 levels. While Colombia and Mexico currently face similar levels of violence, Colombia has achieved a steady decline in homicides, unlike Mexico, which has seen an upward trend over the past decade (Figure \ref{InvestmentByCountry}). 

\begin{figure}[!htbp]
\centering
\includegraphics[width=0.75\textwidth]{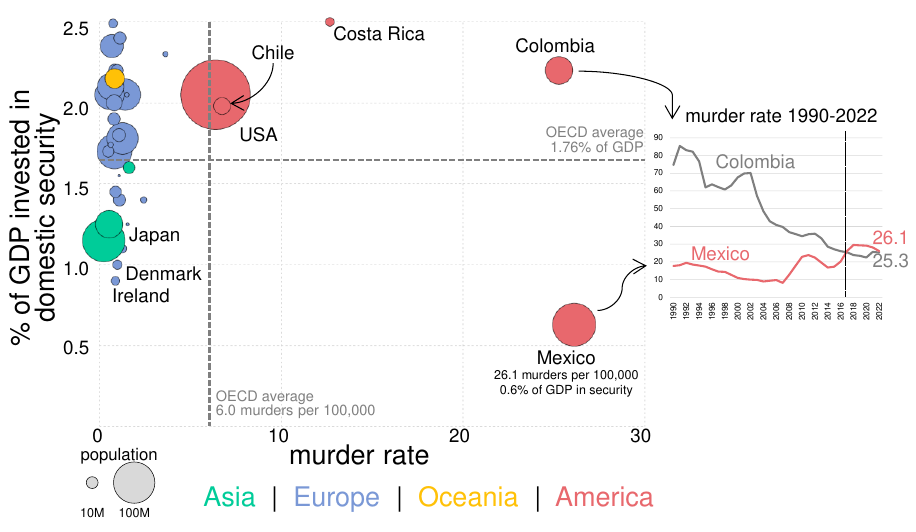}
\caption{Number of homicides per 100,000 people (horizontal axis) and percentage of the GDP invested on domestic security and justice system (vertical axis) by countries. Mexico is the OECD country with the lowest investment in domestic security, followed by Ireland, Denmark, and Finland, countries with less than two murders per 100,000. }\label{InvestmentByCountry}
\end{figure}
}

{
Italy is another country that has experienced substantial reductions in violence \cite{VichiTrendspatternshomicides2020a}. After reaching alarmingly high murder rates in the late 1980s and early 1990s, mainly due to open conflict between the Sicilian mafia and the state, Italian institutions responded forcefully in 1992 following the assassinations of prominent anti-mafia judges Giovanni Falcone and Paolo Borsellino. The Italian state's strategy against organised crime combined the implementation of new special laws, the introduction of high-security prison regimes for mafia members, the creation of dedicated anti-mafia investigative bodies, and the seizure of mafia assets \cite{JamiesonAntimafiaeffortsItaly1998b}. These initiatives proved highly successful and were supported by a substantial increase in public spending on security and law enforcement \cite{PaoliMafiaorganisedcrime2007a}. Today, despite the mafia's continued infiltration into the legitimate economy and political institutions \cite{LeMoglieRevealingMafiaInc2022a, CampedelliMafiaPoliticsMachine2024}, Italy has one of the lowest murder rates in the world. Notably, however, in both Italy and Colombia, anti-crime strategies have primarily focused on security and repression, with less emphasis on designing effective social policies aimed at reducing criminal recruitment and offering legitimate alternatives to vulnerable populations.
}

\section{Methods}

\subsection{Budget for security and social programs}

{
Data from the Federation Expenditure Budget for Fiscal Year 2023 was used to estimate the budget for social programs. It includes programs like \textit{Sembrando Vida} and pensions for young people, mothers, and others in sports and science \cite{PEF2023}. Among all social programs, we consider those that target vulnerable populations to be recruited by a cartel to be the most concerning. These programs include \textit{Apoyo a Madres Trabajadoras, Beca BJ Basico, Beca BJ Medio Superior, Beca BJ Superior, Jóvenes Construyendo Futuro,} and \textit{Sembrando Vida} \cite{PEF2023}. Each year, $\phi = \text{US\$ }8.84$ billion is invested in social programs \cite{PEF2023}, reaching over 10.7 million beneficiaries, and each person receives an average of \text{USD}832. 
}

{
The same data is used to estimate the budget for security programs. It includes the National Guard, Military Forces, Prison System, Justice administration, victim attention and others. The investment made by Mexico in terms of national security programs each year amounts to nearly \text{US\$ }9.88 billion \cite{PEF2023}. This budget includes all security investments which are not directly targeted against one of the cartels. Previous studies related to Mexican cartels estimated that the total cartel population in Mexico is approximately $C = 175,000$ individuals \cite{PrietoCampedelliHope2023}. We split the security budget into three components: $u_1$ and $u_2$, which correspond to the budget used for fighting against $C_1$ and $C_2$, and the remaining part is used against other cartels and other types of crime. We set 80\% of this budget as directly used to fight against organised crime (with 20\% to fight different types of crime) and assume that the budget against organised crime is homogeneously distributed according to cartel size. The top two cartels represent around 14\% of the total cartel size each, so $u_1 = u_2 \approx \text{US\$ }1.11$ billion each year, with a total security budget of $\psi = \text{US\$ }2.22$ billion (Figure \ref{fig:BudgetAppendix}-top).
}

{
\begin{figure}[!htbp]
\centering
\includegraphics[width=0.45\textwidth]{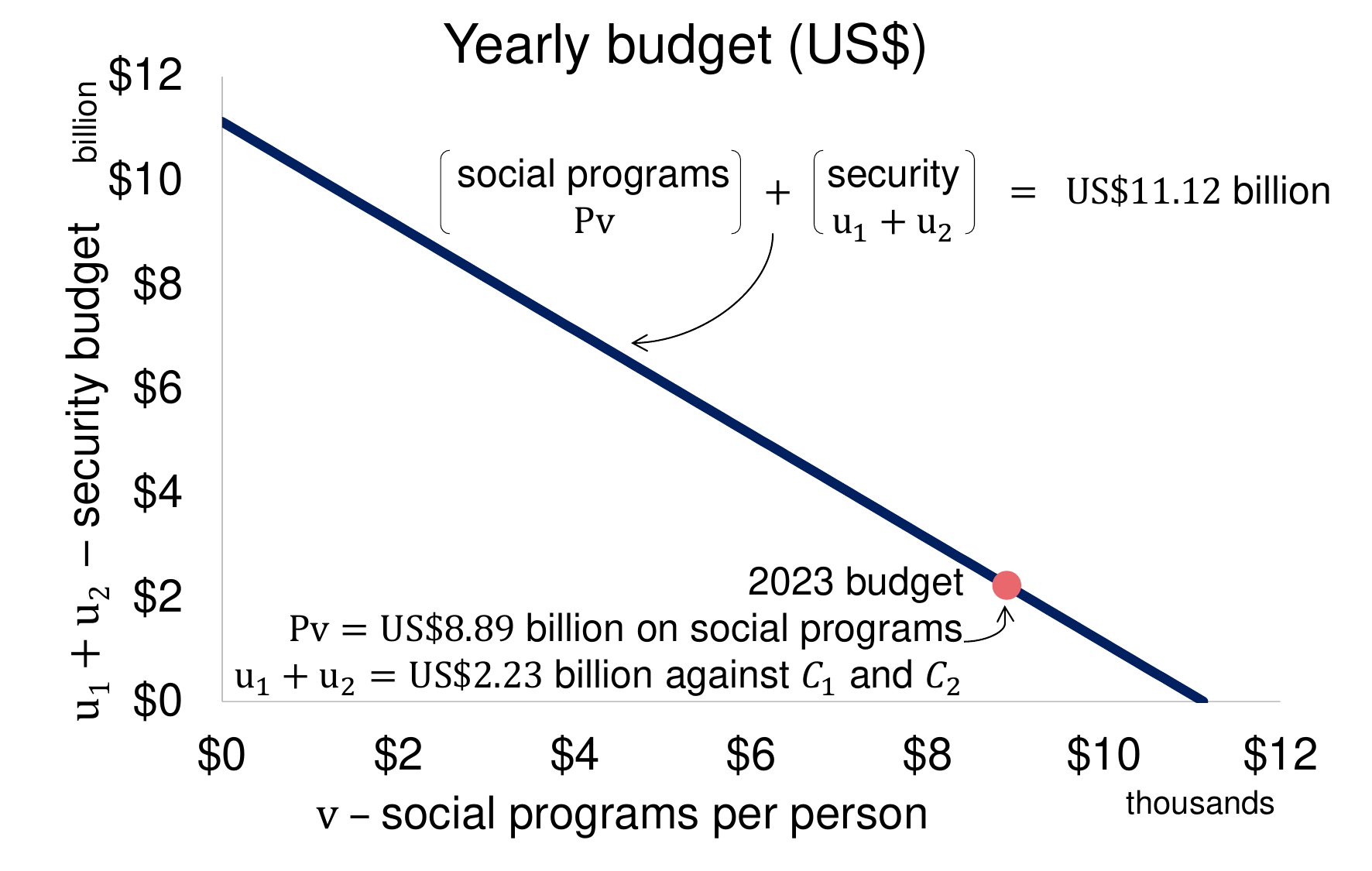}
\includegraphics[width=0.45\textwidth]{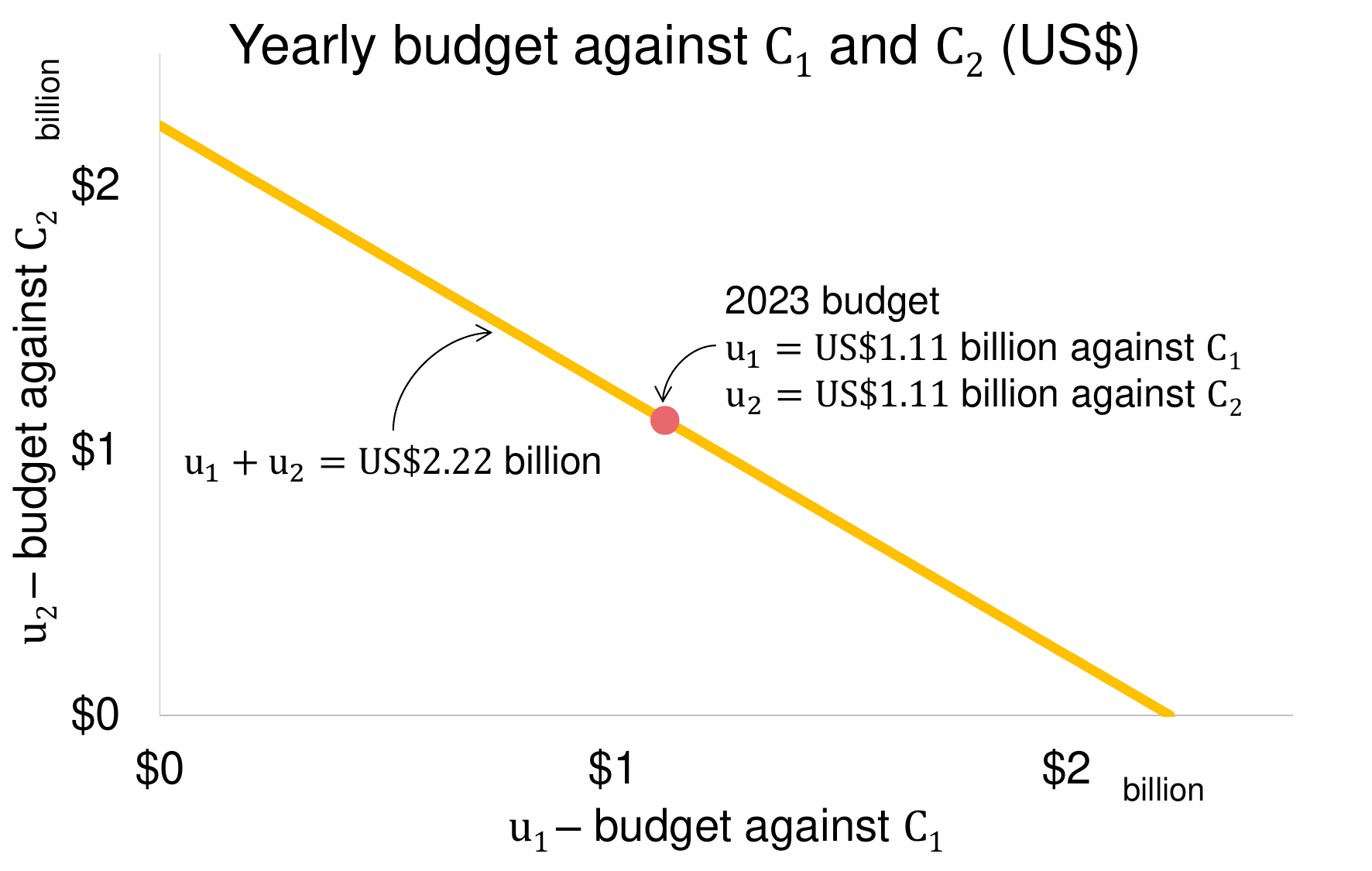}
\includegraphics[width=0.45\textwidth]{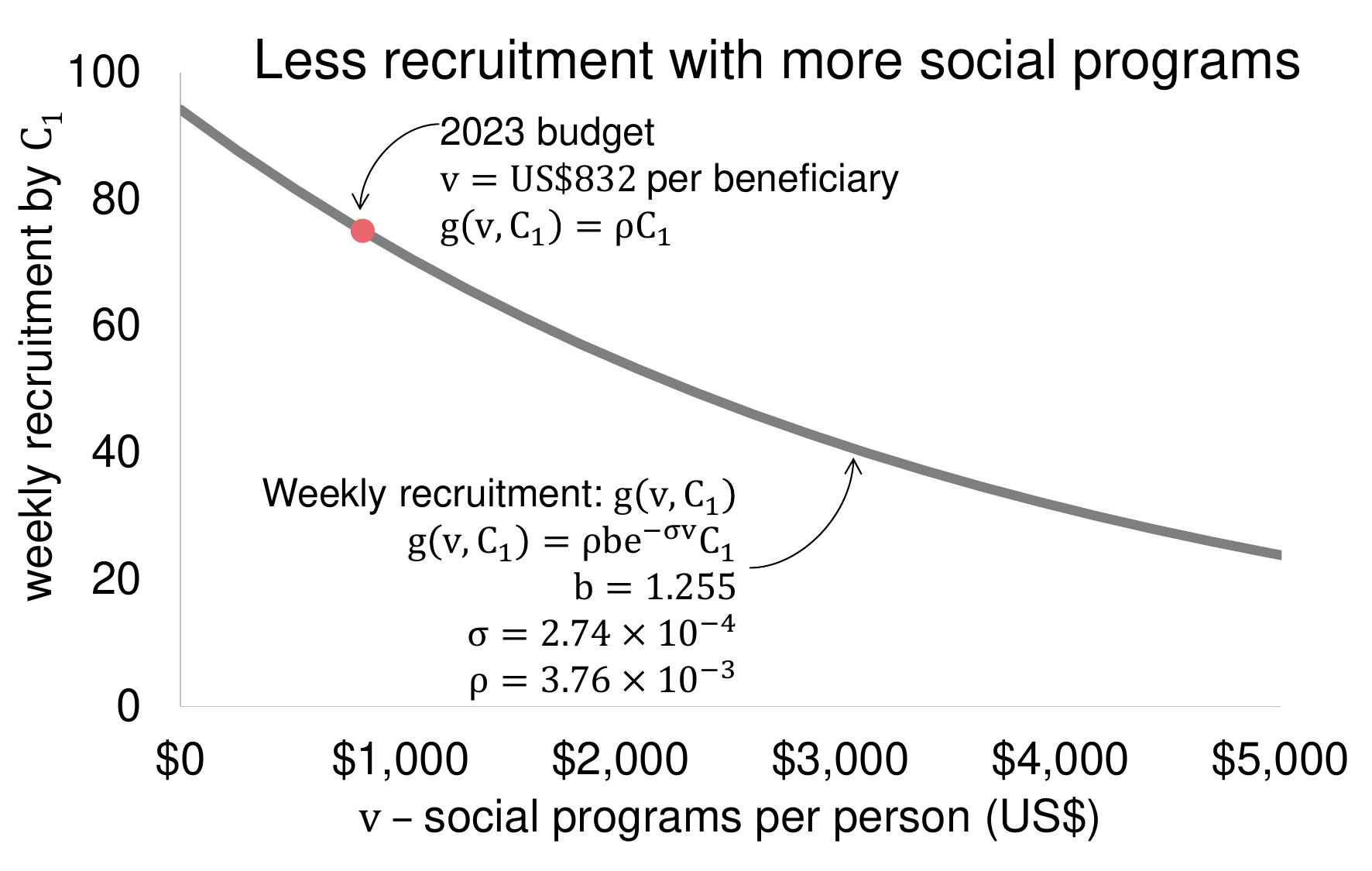}
\includegraphics[width=0.45\textwidth]{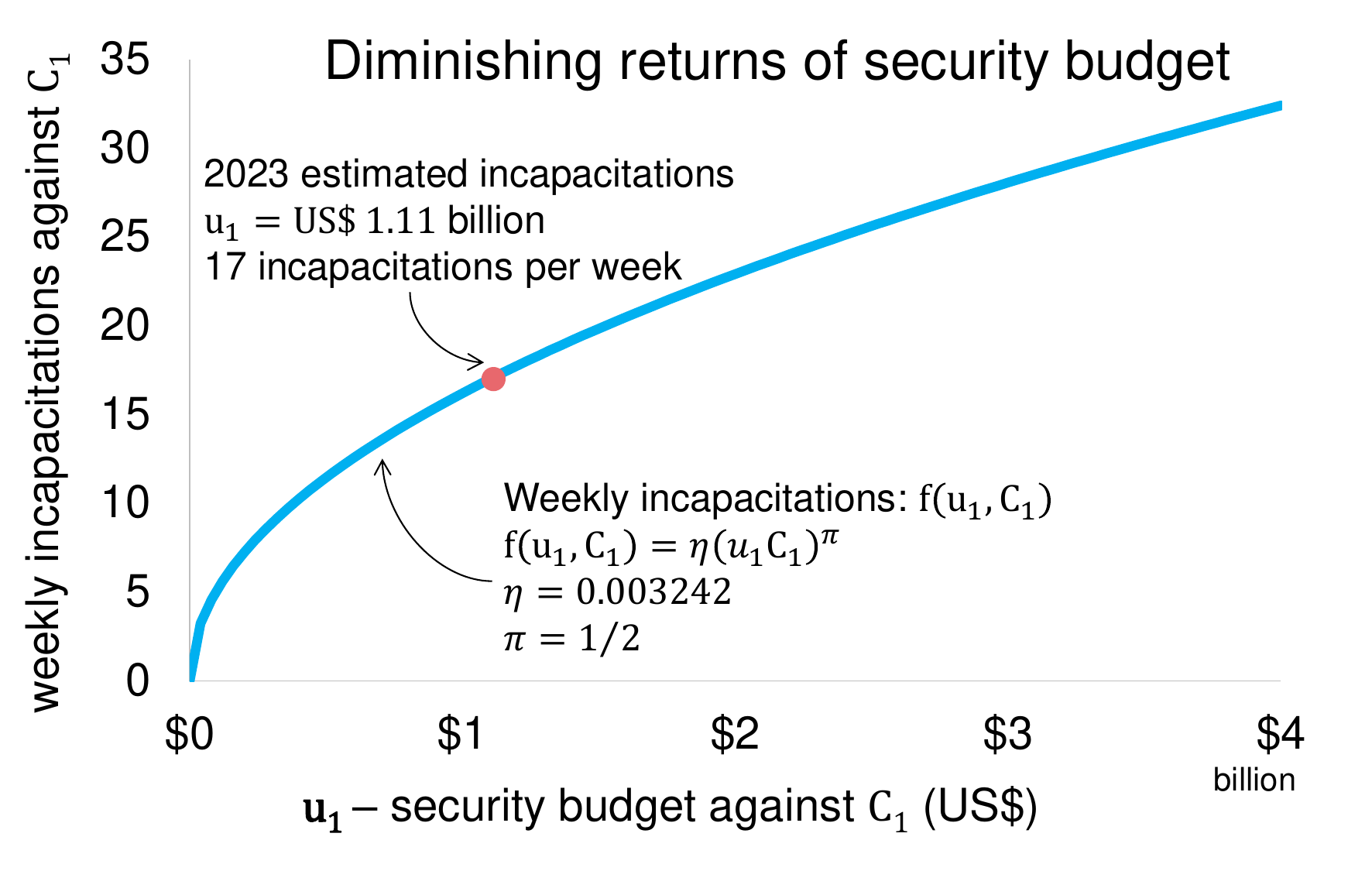}
\caption{The budget for social programs $Pv$ (horizontal axis) and for security programs $u_1 + u_2$ (vertical axis) is, at most, \text{US\$ }11.12 billion (top left). The security budget is distributed among $C_1$ and $C_2$, so $u_1 +u_2= \text{US\$ } 2.23$ billion in 2023 (top right). The diminishing effect of social programs is assumed to reduce cartel recruitment, modelled by an exponential decay (bottom left). The diminishing effect of security programs is to increase incapacitation modelled with an exponent $\pi < 1$ (bottom right). }\label{fig:BudgetAppendix}
\end{figure}
}

{
We assume that it is possible to use some of the security budget for social programs and vice versa. Considering the budget restrictions, we obtain that
\begin{equation} \label{BudgetRestrictionsM}
u_1 +u_2 + Pv \leq \text{US\$ }11.1\text{ billion},
\end{equation}
so the budget can be used for social or security programs (Figure \ref{fig:BudgetAppendix}-bottom). If the entire security budget were allocated to social programs, each person could receive up to \text{USD}1,040 each year. Similarly, if the social budget were used to fight the top two cartels, the yearly amount for fighting against $C_1$ would be $\text{US\$ }5.5$ billion (Figure \ref{fig:BudgetAppendix}-middle).
}

\subsection{Reducing cartel recruitment through social programs}

{
Through social media adverts, it has been observed that a cartel offers roughly twice the income of a formal job and four times the income of an informal job in the capital \cite{bledsoe2025socialmedia, SVyP2024TikTok}. Additionally, cartels not only offer a premium to join, but also often use brutal recruitment methods, including intimidation and coercion, to bring in more members \cite{ForcedRecruitment, jones2018strategic}. With such conditions, it is unclear what the mechanism and magnitude are through which social programs, increased security or even a reduction in poverty prevent cartel recruitment. 
}

{
We construct a cartel participation model taking into account the two types of recruitment (forced and voluntary). We consider that the process occurs as follows. A person who is already part of the cartel finds a potential new member and \textit{exposes} them to the cartel. When a person first meets a member of the cartel, with a probability $\pi_v$ the person is exposed to voluntary recruitment, with probability $\pi_f$ the person is exposed to forced recruitment, and with a probability $\pi_i$ the process fails so the cartel does not recruit the person, so $\pi_v + \pi_f + \pi_i = 1$. Economic models for crime participation suggest that individuals are likely to engage in crime when the expected benefits, accounting for the risk of being caught, surpass the earnings from lawful employment \cite{becker19685, ehrlich1973participation, grogger2000economic}. Here, we model that when a person is exposed to voluntary recruitment, they compare the wage offered by the cartel and the perceived risks of being killed or incapacitated (which we call the ``cartel offer'') with the income from legal activities and social programs (which we call the ``civil offer''). Thus, when a person is exposed to voluntary recruitment, they take into account the wage from the legal market $w$, the value of a social progran $v$, the income from cartel activities $r$, the perceived risk of being incapacitated by state forces $\kappa_P$ and the perceived risk of being killed as part of the cartel duties $\kappa_K$. The perceived risks do not have to be accurate, but reflect how the person feels they will be incapacitated or killed when joining a cartel. When a person is exposed to forced recruitment, they become part of the group regardless of the cartel's premium, social programs or any other conditions. However, when a person is exposed to voluntary recruitment (for example, through social media \cite{bledsoe2025socialmedia, SVyP2024TikTok}), they consider the gains and costs of joining the group. 
}

{
We combine the three scenarios using a mixture model as follows. When a person is exposed to recruitment, they compute the value of 
\begin{eqnarray} \label{linearTheta}
\Sigma  &=& \underbrace{\alpha}_{\text{baseline}}  +\underbrace{ \gamma r -  \delta \kappa_P - \eta \kappa_K }_{\text{cartel offer}} - \underbrace{ ( \gamma w  + \gamma v )}_{\text{civil offer}} \\
&=& \alpha + \gamma (r - w - v) -  \delta \kappa_P - \eta \kappa_K
\end{eqnarray}
where $\Sigma $ is a weighted combination of the perceived risks and yearly income, all in US\$ units, for some values of $\alpha$, which is a baseline parameter, $\gamma$ that is the relative value of income and $\delta$ and $\eta$ reflect how much the person values the risks of being incapacitated or killed. The values of $\delta \kappa_P$ and $\eta \kappa_K$ may be highly heterogeneous in a population. A person might be more risk-averse (higher values of $\delta$ and of $\eta$) or might underestimate the risks of being part of a cartel (lower values of $\kappa_P$ and $\kappa_K$). For more risk-averse individuals, the cartel wage might not compensate for the risks. For a person who underestimates the costs of being part of the cartel, their decision to join is mostly a comparison between the cartel wage and the civil offer. Combining forced and voluntary recruitment, when a person is exposed to a cartel, they join with probability
\begin{equation} \label{CPModel}
P[\text{ join cartel }] = \pi_f + \pi_v \left[ \frac{\exp(\Sigma)}{1+\exp(\Sigma)}\right]. 
\end{equation}
The principle behind this stochastic model is that with a higher civil offer or a higher perceived risk of being killed or incapacitated, the person is less likely to join the cartel (details in the SM - D). 
}

{
Having constructed an individual model for cartel participation, we then observe its collective effects. The civil offer, composed of social programs and legal wages, has two impacts at the collective level (details in the SM - D). First, it reduces the number of people who are exposed to recruitment. With a higher civil offer (so more social programs, for example), fewer people are exposed to a cartel. However, the second impact of a higher civil offer is that many individuals who are exposed to cartel recruitment decide not to join. For example, results of the cartel participation model show that without social programs, more than 100,000 people would be exposed to $C_1$ recruitment in 10 years, and many would decide to join (details in the SM - D). Although offering legal alternatives, opportunities, and social programs should reduce cartel recruitment, some recipients of social support may still engage in criminal activities \cite{chomczynski2023beyond}.  
}

{
Approximately $P = 10.7$ million vulnerable people received a social program in Mexico in 2023. Thus, on average, a person receiving a program obtained $v = \phi/P \approx \text{US\$ }832$ per year. We capture the impact of social programs by a reduction in the recruitment rate of cartels. We express the recruitment of cartel $i$ as $g(v, C_i)$, where the function $g$ captures the diminishing impact of social programs. We capture this reduction by expressing 
\begin{equation} \label{SocialSpendImpact}
g(v, C_i) = \rho b \E^{-\sigma v} C_i,
\end{equation}
where $\rho$ captures the recruitment rate, $b$ is a baseline parameter used for calibrating the estimated recruitment rate of cartels based on the current social programs \cite{PrietoCampedelliHope2023}. The parameter $\sigma>0$ is the \textit{efficiency} of social programs. In 2023, $v = \text{US\$ }832 $ per year, so we write $g(832, C_i) = \rho C_i$, as a baseline. Extant evidence regarding the income received by cartel members is mixed. Based on job listings published to recruit cartel members on social media \cite{CN25000} and on interviews of a former cartel sicaire \cite{CJNG25000}, we take the conservative estimate of $\text{US\$ }13,000$ each year as the income for an average cartel member (roughly 3.5 times the average worker in the country \cite{ENOE2025}). We assume that if a person receives two times what a cartel offers, recruitment will be sporadic (of only $1/1000$). We obtain that $b = 1.255$ and that $\sigma = 2.74 \times 10^{-4}$ (with a range of reasonable values in the SM - G). With these parameters, we are assuming two things: first, if social programs offered the same amount as cartels, then recruitment would only be 3\% of their current rate, and second, if social programs are dropped (with $v=0$), cartel recruitment would increase by 25.5\%. If the budget corresponding to $u_1$ and $u_2$ is used for social programs, then a person could receive $v = \text{US\$ }1,040$ per year, where recruitment would drop $\approx 6\%$ of its current rate. 
}

\subsection{Reducing cartel size through incapacitation}

{
We model the diminishing returns to investments in security as an optimal foraging problem \cite{pyke1984optimal, vandeviver2023foraging}. The principle behind a foraging problem, from the perspective of the security resources, is that they aim to incapacitate as many members of a specific target cartel as possible. When they decide to incapacitate a specific number of people in some city, the result is that some of them belong to the target cartel, but others (perhaps many) do not and are responsible for minor crimes. The intuition behind this is to capture that in Mexico, although many people are incapacitated annually, the majority are for minor offences \cite{ENPOL, ENVIPE2022}. Thus, they have to decide on a strategy for the number of incapacitations among a large number of cities. Through some police intelligence, they can rank locations from the most to the least fraction of the objective cartel. At first, incapacitation is facilitated by the high frequency of cartel members in the most violent locations. Yet, by moving between locations, they encounter fewer cartel members each time, making it more challenging to incapacitate members at a similar rate. Thus, with a ``greedy'' strategy, the security budget has diminishing returns. We capture the diminishing returns of security with the function $f(u_i, C_i)$, expressed as
\begin{equation} \label{SecuritySpendImpact}
f(u_i, C_i) = \eta \left(u_i C_i \right)^\pi,
\end{equation}
where the parameter $\pi$ is the \textit{efficiency} of security programs. We simulate the optimal strategy by considering the presence of $C_1$ members and other criminals across $N$ cities as two independent power-law distributions. Then, we take the optimal strategy when $M$ people are incapacitated (SM - C). The idea is that each incapacitation requires the same security resources, but only a fraction of them (which drops with higher values of $M$) are $C_1$ members. We approximate the diminishing returns with $\pi = 1/2$. It was estimated that the top two cartels suffered 17 incapacitations weekly \cite{PrietoCampedelliHope2023}. With a budget of $u_1(0) =\text{US\$ } 1.1$ billion each year, the parameter becomes $\eta = 0.0232$ (different scenarios in the SM - A to H).
}

{
\paragraph{Cartel shocks in the short and the long run}. Some security policies, such as the beheading of criminal organisations, may have unintended consequences, increasing intra-cartel fighting \cite{calderon2015beheading}. It was observed that most of the effects following the removal of a leader from a cartel (at some time $t_0$) happen within the first six months \cite{calderon2015beheading}. We extend our model by considering a time-varying $\theta(t)$, which captures changes in cartel conflict with the expression
\begin{equation}
\theta(t) = \theta_0(1 + \kappa I_{[t_0, t_0+d]}(t)), 
\end{equation}
where the binary function $I_{[t_0, t_0+d]}(t)=1$ if $t$ is inside the $[t_0, t_0+d]$ interval and zero elsewhere, where $d$ marks a six-month interval after some shock which occurred at time $t_0$, and $\kappa$ captures the magnitude of the shock.  
}

{
For example, with $\kappa = 2$ the number of $C_1$ - $C_2$ homicides triples for six months. Compared against the scenario without a shock, we get that during the year of the shock, the reported homicides increase by more than 90\%. However, since cartels recover after a shock by recruiting new members, the year after the shock, there is a decrease of 3\% in the number of homicides with respect to the baseline (so fewer homicides than before the shock). Combining the costs, the long-term impact of that shock is an increase of only 1.2\% in the 30-year (cumulative) number of homicides (details in the SM - H).
}

{
\paragraph{Cartel fragmentation} The case of cartel fragmentation (when two of its factions divide, often into rival groups \cite{dell2015trafficking}) can be observed in a similar manner. Consider $\theta(t) = \theta_0 I_{[t>t_0]}(t)$, where the binary function $I_{[t>t_0]}(t)=1$ after time $t_0$ when the cartel has fragmented (and the function is zero before $t_0$). Comparing the scenario with fragmentation to the one without, the results show that both factions will have a smaller equilibrium size, thereby reducing the cartel's harm. However, due to fragmentation, there is a sharp increase in cartel homicides, which almost compensate for the drop in cartel harm. The long-run costs of cartels, combining their harm and homicides, drop by approximately 1\% with fragmentation (details in the SM - I).  
}

\subsection{Assigning costs to cartel activities}

{
Violence creates a massive burden in society, not only due to the human lives that are lost, but also the people who are displaced, the additional economic activity that would have happened without violence (such as tourism and investments) and what families and companies must pay to protect from insecurity \cite{FarfanNearshoring}. The direct and indirect costs of violence in Mexico were estimated at \text{US\$ }230 billion in 2022, amounting to 18.3\% of the country's GDP \cite{index2023mexico}. According to the Mexico Peace Index, 45\% of these costs are attributable to homicides, while 55\% are linked to other forms of violence. We analyse these two components separately. In the case of homicides, there were approximately 31,000 intentional killings recorded that year, but many of them are not directly related to cartels \cite{index2023mexico}. We therefore assign an average cost of \text{US\$ }3.3 million per homicide. This estimate, referred to here as the value of a cartel homicide, differs from the conventional value of a statistical life, as it also captures the broader social and economic repercussions of violence, including the enduring impacts on families, neighbourhoods, and communities \cite{de2020value}.
}

{
Secondly, for the non-homicide costs of violence, we apply the Pareto principle, assuming that a disproportionate share of these costs is attributable to criminals affiliated with cartels. We assign $80\%$ for the estimate. We assign the costs uniformly among $C_0 = 175,000$ cartel members \cite{PrietoCampedelliHope2023}. The costs of cartel activities are computed by distributing the impact among the estimated number of cartel members. We obtain a yearly cost of a cartel member of $h = \text{US\$ }0.58$ million. Additionally, we vary the costs assigned to cartels inside the $[70\%, 90\%]$ interval and their size inside the $[160,185]$ thousands interval and get values between \text{US\$ }0.46 million and an upper boundary of \text{US\$ }0.70 million for the yearly cost of a cartel member.
}

\subsection{Optimal budget allocation and comparing budget strategies}

{
We consider different strategies for budget allocation between social programs or fighting the top two cartels. For a strategy that is within budget, we assume that $u_1(t) + u_2(t) + P v(t) \leq \text{US\$ } 11.1$ billion for all values of $t$. By varying the security and the social budget, different recruitment and incapacitation rates are obtained in Equation \ref{eq:MasterEquation}. Further, we also consider the case with an unlimited budget (SM - D). 
}

{
We compare strategies by considering the controls $T = \left\{ u_1(t), u_2(t), v(t) \right\}$ and estimating the long-term costs $Q(T)$. The total cost of the current budget allocation is given by $Q(u_1 = 1.1, u_2 = 1.1, v = 832) \approx \text{US\$ } 804$ billion (SM - I). The problem in Equation \ref{ObjectiveEq}, with dynamics from Equation \ref{eq:MasterEquation}, and with budget restrictions from Equation \ref{BudgetRestrictionsM}, is an optimal control problem with an infinite time horizon (SM - F). It can be solved with the standard Maximum Principle, giving a set of necessary optimality conditions for optimality \cite{grass_optimal_2008}. The numerical solution is obtained using Matlab \cite{MATLAB} with code available \href{https://github.com/rafaelprietocuriel/ControlTheoryOptimizationOC}{in this repository.}. The worst strategy is given by $T_0 =\left\{ 0,0,0 \right\}$, with $Q(T_0) = \text{US\$ }933$ billion. Results show that the optimal strategy is an unlimited and dynamic budget, with $Q(T^\star) = \text{US\$ }821$ billion. 
}

\subsection{Modelling cartel's profit}

{
To detect how a cartel reacts to more social or security programs, we model their income, costs and profit (details in the SM - G). We take their income $Y(N) = \alpha N^\beta$ to be a sublinear function (so $\beta \in (0,1)$) with respect to the number of members $N$ (or a Cobb–Douglas function taking into account only labour) and take their costs $X(N) = (r + \omega) N$ to be linear, depending on the average salary they pay $r$ and additional costs per person $\omega$. For different parts of their activities, such as drug trade, extortion or kidnapping, increasing income is not a matter of just a few simple steps; otherwise, they would have already taken those steps. Thus, we assume that cartels mostly adjust their costs (so adjust their size) as their optimal strategy (details in the SM - G).
}

{
If cartels follow an optimal strategy, then social programs have three impacts. First, recruiting and maintaining members becomes more expensive, so the optimal size is smaller. With a smaller and more costly cartel, their profit is also smaller, and they create less harm and fewer casualties (SM - G). 
}

\subsection{More efficient social programs or more efficient security programs}

{
To detect the impact of the efficiency of social programs, we vary the premium salary offered by cartels by $\pm 20\%$. With these two cases, we consider the set in which $b_m = 1.2075$ and $\sigma_m = 2.27 \times 10^{-4}$ (where the social programs have a smaller impact) and the set with $b_M = 1.3321$ and $\sigma_M = 3.29 \times 10^{-4}$ (where social programs have a more significant effect). We then analyse the current budget allocation, as well as the scenario where social and security programs are dropped and the optimal scenario (SM - G and H). 
}

{
For the efficiency of security programs, we analyse the results with $\pi \pm 20\%$, meaning that we analyse what happens if the police are more or less efficient at targeting a specific cartel with its resources (SM - H). To investigate how sensitive the results are to the lethality of cartels, we analyse the weekly number of homicides with $\pm 20\%$. To examine how sensitive the results are in terms of having asymmetric cartels, we analyse the base case in which $C_1(0) = C_2(0) = 25,000$ and an asymmetric case, in which $C_1(0) = 31,500$ and where $C_2(0) = 18,500$. 
}

\subsection{Sensitivity analysis}

{
\paragraph{Sensitivity tests and summary of results.} Our model rests on a few assumed parameters. For this reason, we conducted extensive tests to assess the range of outcome variation and ensure that our assumptions do not solely drive our main results (SM - E). To determine the degree to which the results are sensitive to the parameters and assumptions made, we analyse the current budget assignment and two additional scenarios: one without social and security programs and the other, the optimal strategy without budget restrictions. We then vary the parameters elected within a reasonable range and detect whether the critical results hold.
}

{
Results show that varying the value of the parameters within a reasonable range has a minor impact on the quantitative description of costs (details in the SM - D, E, G and H). Although there is an impact on the specific amounts, there is no impact on the description and the implications. In other words, it is worth investing in social and security programs, but currently, they are not enough to reduce violence and cartel harm. Also, in the long run, it is convenient to invest more resources in social and security programs, regardless of whether the efficiency of social programs to stop cartel recruitment varies $\pm 20\%$ or whether social programs are more efficient or whether the initial size of cartels is not symmetric.
}

\renewcommand{\figurename}{Supplementary Figure}
\renewcommand{\tablename}{Supplementary Table}

\section*{Appendix}

\subsection*{A - Parameters of the model}

{
Previously, it has been estimated that cartels have roughly 175,000 members \cite{PrietoCampedelliHope2023}. This number was obtained by estimating the increase in the number of members due to recruitment and the decrease related to incapacitations, homicides, missing people and retirement. Official data from the number of homicides in the country \cite{InegiMortalidad}, the number of people missing and yet to be found \cite{RNPDNO}, and the Mexican prison census \cite{inegiPenitenciario} were used to obtain those estimates. We analyse the size of each cartel $C_i(t)$ as it varies in time $t$ measured in weeks and consider four reasons why it varies \cite{PrietoCampedelliHope2023}. We consider that 
\begin{eqnarray}
\dot{C}_i (t) & = & \underbrace{g(v(t), C_i (t))}_\text{recruitment} - \underbrace{f(u_i (t), C_i (t))}_\text{incapacitation} - \underbrace{\theta C_i (t) C_j (t)}_\text{conflict} - \underbrace{\omega C_i^2 (t)}_\text{saturation}, \qquad i=1,2. 
\end{eqnarray}
We model the impact of social programs with $g(v, C_i) = \rho b \E^{-\sigma v} C_i$, which depends on the \textit{efficiency} of social programs $\sigma$, the recruitment rate $\rho$ and a baseline parameter $b$ which is set so that with the current value of social programs, $g(v_0, C_i)= \rho C_i$. The impact of security programs is modelled with $f(u_i, C_i)= \eta \left(u_i C_i \right)^\pi$, capturing how the budget $\text{US\$ }u_i$ used for fighting cartel $i$ via security expenditures by the state, depending on the \textit{efficiency} of security programs, $\pi$, and a baseline parameter $\eta$. The parameter $\theta$ models the homicides that $C_i$ inflicts on $C_j$. The parameter $\omega$ captures the rate at which members of a cartel retire due to internal instability, e.g., group dropouts, disputes, retirement and so on. The reasoning for picking each function and its corresponding parameters is outlined below.
}

\subsubsection*{Social programs} 

{
Social programs are often portrayed as part of the government's strategy to deter young men from joining criminal gangs \cite{economist2025mexico}.  Although the core objective of social programs is not directly to reduce crime participation, they are also part of the broader strategy for preventing violence \cite{martinez2023politica}. Social programs have recently become universal, largely unconditional, and are often delivered as cash handouts \cite{economist2025mexico}. Additionally, cash transfers have expanded rapidly, often at the expense of other social programs that addressed specific needs, such as healthcare access, and targeted particular regions \cite{economist2025mexico}.
}

{
We assume that social programs have an impact on cartel recruitment. Models for crime participation usually assume that a person will commit crimes (or join a criminal group) if the returns to commit crime exceed the gains of other (licit) activities, considering the risks of apprehension and the individual costs of crime participation \cite{grogger2000economic}. Therefore, we theorise that a cartel must offer a premium income for recruiting members (compensating for the risks of incapacitation and injuries for cartel members). Therefore, with higher values of a social program, there exist fewer relative gains and, thus, motivations to join cartels \cite{ehrlich1973participation, freeman1999economics}. We model this by assuming that with higher values of a social program $v$, cartel recruitment becomes more challenging (thus, the function $g$ decreases). We write $v$ as the value that each person receives each year and express cartel recruitment as
\begin{equation}
   g(v, C_i)  = \rho b \E^{-\sigma v} C_i,
\end{equation}
where the component $b \E^{-\sigma v}$ captures the impact of social programs. We take $b \E^{-\sigma v} = 1$ for the current levels of spending $v_0 = \text{US\$ }832$ per year. We estimate that the income for an average cartel member is $w_0 = \text{US\$ }13,000$ each year, based on job listings published to recruit cartel members on social media and interviews of a former cartel sicaire \cite{CN25000, CJNG25000}. We assume that if social programs double the cartel income, recruitment would be rare (only $1/1000$). Thus, we take $b \E^{-2 w_0 \sigma} = 1/1000$ to capture the impact of a high value of social programs. With those two points, we get that $b = 1.255$ and that $\sigma = 2.74 \times 10^{-4}$. For cartel $C_1$, it was estimated weekly recruitment of 75 members \cite{PrietoCampedelliHope2023}. Thus, we obtain that $\rho = 3\times 10^{-3}$.
}

\subsubsection*{Conflict between cartls} 

{
For the lethality of the conflict between cartels, it was estimated that any of the top two suffer 7.5 casualties each week \cite{PrietoCampedelliHope2023}. This results in $\theta = 1.2 \times 10^{-8}$, which means that we estimate the value of $s \theta = 0.0796$ to the cost of the homicides between $C_1$ and $C_2$. With this value and the size of $C_1$ and $C_2$, each cartel suffers 7.5 casualties each week, and there are about 15 cartel casualties combined. The number of casualties of $C_1 = \theta C_1 C_2 \approx 7.5$ each week. The total number of homicides between then is $ 2 \theta C_1 C_2 \approx 15$ each week.
}

{
We assume that without any external forces, cartels have a maximum size of $C_{Max} = 50,000$ individuals, resulting from saturation effects. Based on that, we then estimate that the parameter $\omega = \rho / C_{Max} = 5.2 \times 10^{-8}$. Thus, the saturation parameter is $\omega = 5.2\times 10^{-8}$.
}

\subsubsection*{Objective function} 

{
Combining the four costs, we get that the objective is to reduce the costs of social budget, security, and the impact of cartel size and homicides, so we write
\begin{equation} 
Q: \int_0^\infty \E^{-rt} \left( \underbrace{2s \theta C_1(t) C_2(t)}_{\text{homicides}} + \underbrace{ h ( C_1(t) + C_2(t))}_{\text{cartel harm}}+  \underbrace{ u_1(t) + u_2(t)}_{\text{security}}  + \underbrace{P v(t)}_{\text{social}} \right) \,\Dt,
\end{equation}
where $s$ is the social cost of a cartel member being murdered by another cartel, whilst $h$ encapsulates the cost of cartel harm. A weekly discount rate of $r = 0.001$ is considered. 
}

\subsubsection*{Costs of cartel harm, homicides and controls} 

{
It is possible to estimate the monetary value of saving a youth by considering the value of statistical life \cite{cohen1998monetary}. This concept has been used to compare the costs of offenders over time \cite{delisi2010murder}. Yet, that value assigns a similar cost to the victim of a homicide than to the victim of any other cause of death. However, homicides create many more indirect costs from the fear they create. Thus, instead of taking the costs of homicides directly, we use the estimates of the cost of violence. Considering the direct and indirect costs of violence, it was estimated to be a total of \text{US\$ }230 billion \cite{index2023mexico}. Of the costs of violence, the Mexico Peace Index estimates that 45\% is related to homicides and 55\% are other types of violence and indirect costs. To determine the financial cost of a homicide, we take the estimated monetary cost of homicides in 2023 ($\text{US\$ }102.8$ billion) and distribute the costs homogeneously across the 31,000 homicides in the year. The cost of each homicide is $\text{US\$ }3.3$ million. The conflict between $C_1$ and $C_2$ results in roughly 15 homicides each week, so we get that the parameter $s = 0.0796$.
}

{
To determine the financial cost of a cartel member, we assume that they create most violence, so we assign $80\%$ for the estimate. We assign the costs uniformly among $C_0 = 175,000$ cartel members \cite{PrietoCampedelliHope2023}. The yearly burden of cartel activities is computed by distributing the impact among the estimated number of cartel members. With an estimate of 175,000 cartel members, we obtain a yearly cost of a cartel member of $h = \text{US\$ }0.58$ million. We vary the assignment to cartels inside the $[70\%, 90\%]$ interval and the size of cartels inside the $[160,185]$ thousands interval to quantify the impact of the parameters. We obtain that a cartel member costs between $\text{US\$ }0.50$ million and $\text{US\$ }0.69$ million each year. 
}

{
In the system of equations, there are three control variables, given by:
\begin{itemize}
\item $u_1$ - yearly budget against $C_1$. With $u_1(0) = \text{US\$ }1.11$ billion

\item $u_2$ - yearly budget against $C_2$. With $u_2(0) = \text{US\$ }1.11$ billion

\item $v$ - value of social programs, with $v(0) =  \text{US\$ }832$ per year.
\end{itemize}
}

{
The budget restrictions are
\begin{equation}
Pv + u_1 + u_2 = \text{US\$ }11.1\text{ billion each year,}
\end{equation}
where $P$ is the number of beneficiaries (10.7 million). 
}

\clearpage

\subsubsection*{Parameters of the model} 

{
\begin{table}
    \centering
    \begin{tabular}{lccl}
        Term & symbol & value \\
        \hline
        lethality & $\theta$ & $1.2 \times 10^{-8}$ \\
        saturation & $\omega$ & $7.53 \times 10^{-8}$\\
         impact of social budget & $g(v, C_i)$ & $\rho b \E^{-\sigma v} C_i$ \\
         base recruitment& $b$ & $1.255$ \\
         recruitment rate& $\rho$ & $3\times 10^{-3}$\\
         social spending efficiency& $\sigma$  & $2.7 \times 10^{-4}$ \\
         impact of security budget & $f(u_i, C_i)$&$\eta (u_iC_i)^\pi$, $\eta = 0.02322$ \\
         efficiency of security spending & $\pi$ & $1/2$ \\
         cartel-cartel homicides parameter & $s$ & 0.0796\\
         yearly cost of a cartel member & $h$ & \text{US\$ 0.58} million \\
         beneficiaries & $P$ & 10.7 million \\
         weekly discount rate & $r$ & 0.001  \\
         \hline
         yearly social budget & $\phi$ & \text{US\$ 8.9} billion \\
         social program per person & $v$ & $\text{US\$ 832}$ \\
         budget against $C_1$ & $u_1$ & \text{US\$ 1.11} billion\\
         budget against $C_2$ & $u_2$ & \text{US\$ 1.11} billion \\
         \hline
    \end{tabular}
    \caption{Parameters of the model}
    \label{tab:my_label}
\end{table}
}

\clearpage

\subsection*{B - The cartel ecosystem}

{
The data used to estimate cartel size in \cite{PrietoCampedelliHope2023} is from 2020. However, the cartel landscape has undergone significant shifts over the past five years. Firstly, CJNG has expanded significantly in recent years, with a higher presence in many states in the central and southern parts of Mexico. 
}

{
The second largest cartel was the Sinaloa Cartel. However, following the kidnapping and imprisonment of Mayo Zambada in July 2024, the cartel has fragmented into two factions: Chapitos and La Mayiza. 
}

{
Additionally, in February 2025, the State Department designated the Sinaloa Cartel, the CJNG and other Latin American criminal groups as global terrorist organizations. There are many changes and shocks that are difficult to capture on a system of equations, where many parameters are also modelled. Consequently, our model assumes two major cartels of equal size, though the underlying dynamics are more complex. We also introduce variations in the initial size within SM - D.
}

\subsection*{C - Diminishing security returns}

{
We capture the returns to investing more resources in security based on a principle from optimal foraging \cite{pyke1984optimal, vandeviver2023foraging}. The principle of optimal foraging was conceived based on how an animal tries to find optimal places for feeding. The idea is that an animal encounters a patch for foraging. Once it finds a patch, it preys (or eats) there. The food in that patch is reduced because of that animal, so it needs to move and find a different patch. An optimal strategy for the animal is to move to a different patch when the patch produces the mean, taking into account the time required for the move. The same idea has been applied to crime, where criminals prey on various locations where they can find suitable targets, and the patches correspond to the types of victims they will target (such as houses in the case of burglary) \cite{sorg2017explaining, johnson2014offenders}. We model a policing strategy based on the same principle of optimal foraging. Here, each patch corresponds to a different city in the country. The approach followed by the police is based on deciding how many incapacitations will be performed in each city \cite{vandeviver2023foraging, sorg2017explaining}. Each incapacitation may correspond to members of the objective cartel or may correspond to members of other cartels or other criminals not affiliated with a cartel. The ``gain'' of the strategy is the effective number of members of $C_1$ who are incapacitated. 
}

{
Formally, we model the distribution of cartel members in city $i$ as a combination of members of $C_1$ and other criminals (perhaps ones that are not affiliated with any cartel and are responsible for other types of crime, such as robberies). We consider $c_i$ to be the number of members of $C_1$ in the city $i$ and $y_i$ the other criminals in the city (so the city $i$ has $c_i + y_i$ criminals). The distributions of $c_i$ and $y_i$ are not homogeneous across all cities and are uncorrelated. In some cities, most of the criminals are members of $C_1$, but in others, there are many members of different cartels, and finally, there are some cities with only a few cartel members (Figure \ref{ForagingResults}-A). 

\begin{figure}[!htbp]
\centering
\includegraphics[width=0.65\textwidth]{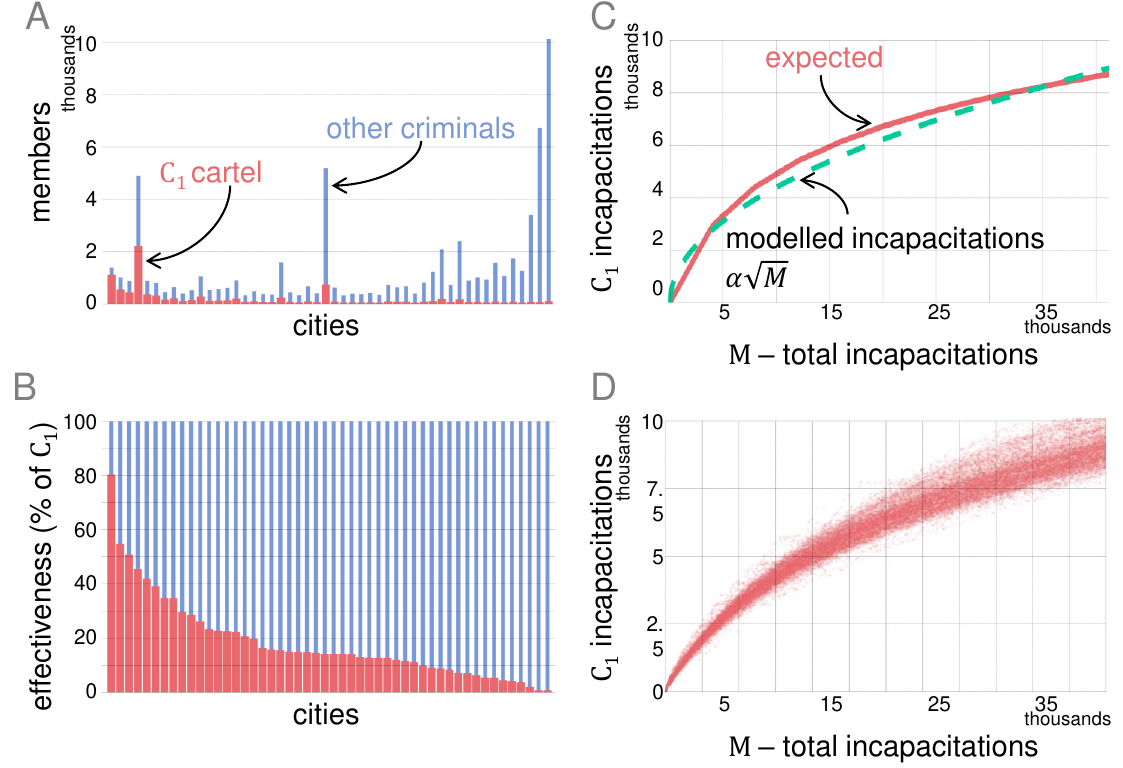}
\caption{A - Number of $C_1$ members (red) and $Y$ members (blue) across different cities (top). Cities are sorted from the one with the highest fraction of $C_1$ members, expressed as $c_1$, to the one with the lowest share of $C_1$. B - Share of $C_1$ members across $N$ cities. C - Expected and modelled number of $C_1$ incapacitations (vertical axis) as a function of the total incapacitations by the police (horizontal axis). D - Number of $C_1$ incapacitations (vertical axis) when the police incapacitated a given number of individuals (horizontal axis) with a varying distribution of cartel members across cities. Each dot is the result of one assignment of $C_1$ and $Y$ members with $M$ incapacitations. The process is then repeated 500 times, obtaining possible departures from the modelled returns. }\label{ForagingResults}
\end{figure}
}

{
We assume that security resources are distributed with the objective of maximising their impact. Thus, out of $N$ cities, the Police will distribute their resources so that they can hinder the highest number of $C_1$ members. The optimal strategy for the police is to distribute its resources across the $N$ cities to maximise the expected number of $C_1$ members that can be incapacitated. Thus, the Police distribute their resources across cities, incapacitating $M_i$ individuals in city $i$, with $M = \sum_i M_i$ being the number of people targeted. Across all cities, and due to the secretive nature of cartels, the police target members of different cartels and other criminals, say $Y_i$. We model the \textit{effective} number of $C_1$ members that are incapacitated in city $i$ as a hypergeometric distribution. The hypergeometric distribution is a discrete distribution which considers the probability of $k$ successes (incapacitated members of $C_1$) in $M_i$ draws (or $M_i$ incapacitated people in city $i$), without replacement, from a finite population of size $x_i = c_i + y_i$ (the number of criminals in city $i$), where $c_i$ of those criminals are part of $C_1$. Formally, the expected number of people incapacitated from $C_1$ in city $i$ is
\begin{equation}
P\left[ \text{incapacitating $k$ members of $C_1$ in city $i|M_i$}\right] =  \frac{\binom{c_i}{k} \binom{x_i - c_i}{M_i - c_i}}{ \binom{x_i}{M_i}}.
\end{equation}
The expected number of $C_1$ members incapacitated in city $i$ is then $M_i (c_i / x_i)$, so the coefficient $c_i/x_i$ captures the rate of incapacitations of $C_1$ elements in city $i$ (where $M_i (y_i / x_i)$ is the ineffective incapacitations). 
}

{
An optimal strategy, considering the effective number of incapacitations across cities, is to sort cities in decreasing order based on the ratio $c_i / x_i$ and target those with the highest rate. Thus, it is possible to sort cities based on $c_i/x_i$ in decreasing order (Figure \ref{ForagingResults}-B). However, the city with the highest ratio $c_i/x_i$ is not necessarily the one with the largest number of members of $C_1$, but the one with the highest share. With $M=1$ (so a single incapacitation), the best strategy is to target the individual from city $1$, as it maximises the expected number of $C_1$ incapacitations. The probability of success (the person being from $C_1$) is $c_1/x_1$. If it was successful (so the incapacitation was from $C_1$), then a second incapacitation in that city has a probability $(c_1-1)/(x_1-1)$ of success (which is slightly smaller than the initial probability $c_1/x_1$). The optimal strategy will continue with $q_1$ criminals in the city $1$ (of which $p$ criminals were part of $C_1$), until the ratio $(c_1 -p_1)/(x_1-q_1+p_1)$ is equal to $c_2/x_2$. After $q_1$ incapacitations, the police are indifferent in targeting cities $1$ and $2$ as the expected number of $C_1$ members is the same. A similar process will happen after $q_2$ incapacitations between the top two cities. After $q_2/2 + q_1$ incapacitations in city $1$ and $q_2/2$ in city $2$, the police are now indifferent between the top three cities. The same process occurs when moving between the top cities. Subsequently, the probability of incapacitating $C_1$ members decays as more of their members are incapacitated and the police resources are distributed among more cities. Thus, there are diminishing marginal returns to incapacitation, even if the police are capable of detecting the city with the highest presence of the cartel they aim to target, and they follow an optimal strategy. Each time the police move to a new city, they encounter a smaller ratio $c_i/x_i$, becoming, across each step, less efficient.
}

{
To test how rapidly the decrease in efficiency might be, we assume a distribution of $C_1$ members across cities and an independent distribution of $Y$ members. In both cases, we consider that the distribution of members across $N$ cities follows a power-law distribution. Thus, we expressfor city $j$ the number of $C_1$ members as $c_j = C_1 j^\beta/\sum_{j = 1}^N j^{\beta}$. We take $\beta = -1$, meaning that in the city with the highest number of $C_1$ members, there will be approximately $C_1 / \sum_{j = 1}^N j^{\beta}$ members of $C_1$. For the members of $Y$ in city $k$, we also consider an independent power-law distribution, so $y_k = Y k^\beta/\sum_{k = 1}^N k^{\beta}$, with the same value of $\beta$, so that $Y = \sum_{k=1}^N k^{\beta}$. To mimic the number of cartel members across cities in Mexico, we consider $N = 50$ cities, with an estimated size of $C_1 = 25,000$ cartel members \cite{PrietoCampedelliHope2023}. Then, we take $Y = 250,000$ for members of other cartels and other criminals in the country. The police decide then to target $M$ individuals and distribute the incapacitations across cities according to the optimal strategy (Figure \ref{ForagingResults}-C).  
}

{
We repeat the same process of taking a distribution of members of $C_1$ and an independent distribution of $Y$ members across $N$ cities. For each repetition, we randomly alter the distribution of members of $C_1$ and $Y$. Then, for each assignment, we consider the optimal strategy to follow given the distribution of $C_1$ and $Y$ (Figure \ref{ForagingResults}-D). Then, we compute the expected number of effective incapacitations of $C_1$ members with $M$ incapacitations. By repeating the same process a sufficiently large number of times (500 in our case), we consider the possible departures that could be observed under a varying distribution of cartel members across cities. 
}

{
We approximate the number of $C_1$ incapacitations by the expression $\alpha M^\pi$. Although there are some departures from the expected number of incapacitations, we obtain that the diminishing returns of increasing the number of incapacitations can be captured by an exponent smaller than one. We capture the diminishing returns of security programs, with the number of incapacitations following
\begin{equation}
    f(u, C_1) = \eta (u_1 C_1)^\pi \text{, with $\pi < 1$,}
\end{equation}
where the coefficient $\pi=1/2$ captures diminishing returns. 
}

{
For the parameter estimation, we consider that in Mexico, it was estimated that cartels suffer 110 incapacitations each week, roughly 17 of them corresponding to the biggest cartel \cite{PrietoCampedelliHope2023}. With a yearly budget of\text{US\$ }1.11 billion, we get that
\[ \eta ( u_1 C_1)^\pi = 17,\]
so with $\pi = 1/2$ we obtain that $\eta = 0.02322$. 
}

\subsection*{D - A cartel participation model} 

{
Why would a person join a cartel, considering the risks they will be exposed to of being killed by a rival group or being incapacitated by the state? What does the cartel offer that maintaining legal activities does not? In some parts of Mexico City, for example, the formal sector pays on average $\text{US\$ }5400$ per year, while informal jobs pay only $\text{US\$ }2700$. In contrast, cartel recruitment ads, often found in social media, offer up to $\text{US\$ }11000$ with possible increases depending on ``performance'' \cite{bledsoe2025socialmedia}. Cartels are promising recruits \textit{buena paga semanal, hospedaje y comida, vacaciones, equipo táctico, calzado y ropa}, meaning ``good weekly pay, lodging and meals, vacations, tactical gear, footwear, and clothing'', highlighting the financial premium of joining organised crime \cite{SVyP2024TikTok}.
} 

{
If the cartel offers twice the income of a formal job and four times the income of an informal job, what other sources of income (such as social programs) could be offered to balance the premium of cartels? And would they effectively reduce recruitment? Although offering legal alternatives, opportunities, and social programs should reduce cartel recruitment, some recipients of social support may still engage in criminal activities \cite{chomczynski2023beyond}. Additionally, cartels not only offer a premium to join, but also often use brutal recruitment methods, including intimidation and coercion, to bring in more members \cite{ForcedRecruitment, jones2018strategic}. Thus, it is unclear what the mechanism and magnitude are through which social programs reduce cartel recruitment. Our model for cartel participation (as described in the Methods) captures how a person compares the civil offer and the cartel offer and modifies their probability of joining a cartel accordingly. 
}

{
There are heterogeneous outcomes of the recruitment process. Two individuals with the same perceived risks and weights, who are exposed to the same cartel offer, may make different decisions regarding whether to join a cartel or not. The stochastic model for cartel participation captures the impact of certain elements that an individual considers when deciding to join a cartel. A similar expression can be obtained with a threshold model for collective behaviour \cite{granovetter1978threshold}. For that type of model, the stochastic component hinges on the amount required to persuade a person to join the group \cite{braun1995individual}. Recruitment and incapacitation often involve nonlinearities and threshold effects, where small changes in incentives or pressure can trigger disproportionate reactions. For example, for a small risk of being killed by another cartel, any increase has a negligible impact on cartel recruitment as other elements compensate for that risk. For a high risk of being killed, people mostly join a cartel only by force, but not voluntarily, since the dangers of being killed outweigh the offered wage. Similarly, for high salaries from the legal market or values of a social program, the cartel offer will be too small, and most people would not join (Figure \ref{IndividualImpacts}).

\begin{figure}[!htbp]
\centering
\includegraphics[width=0.95\textwidth]{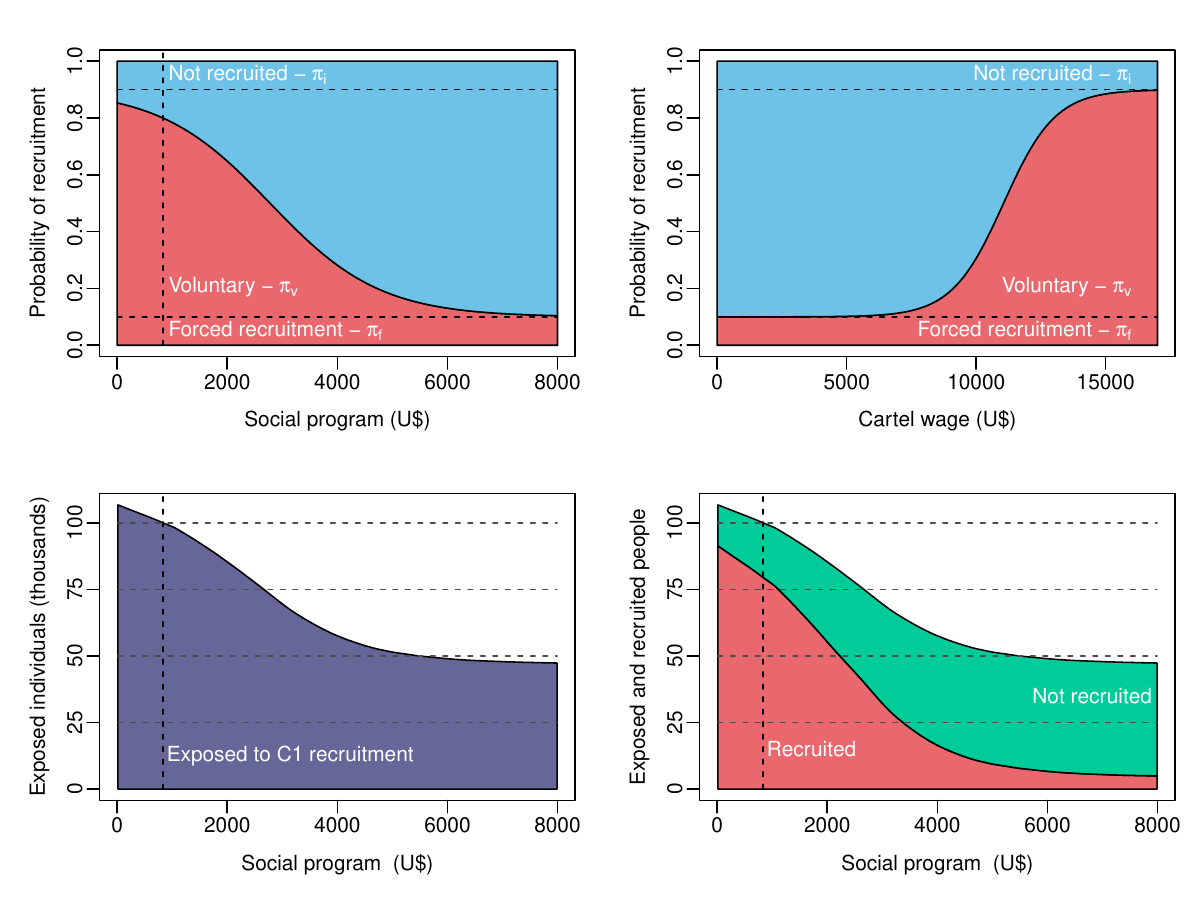}
\caption{Top left - Modelled probability of joining a cartel (vertical) depending on the value of a social program (horizontal axis). To approximate the impact of social programs on the individual's probability of being recruited by a cartel, we use the baseline values of all parameters and vary the social programs. The vertical dashed line corresponds to the current value of social programs. Top right - Modelled probability of joining a cartel (vertical) depending on the cartel wage (horizontal). For a small cartel wage, forced recruitment occurs, but with a higher wage, voluntary recruitment becomes more prevalent. Bottom left - Number of people being exposed to $C_1$ recruitment (vertical axis) depending on the value of social programs $v$ (horizontal axis) for 10 years. The vertical line corresponds to the current value of social programs and the parameters used for the model are $\pi_f = 0.1$, $\pi_v = 0.8$, $\delta = 5000$, $\eta = 10,000$, $\gamma = 1/1000$ and the other parameters correspond to the estimated incapacitation rate, lethality, income of cartels \cite{PrietoCampedelliHope2023}. Bottom right - Of all individuals who are exposed to cartel recruitment, we divide them between those who decide to join (in red) and those who decide not to join a cartel (vertical axis), varying the social programs (horizontal). }\label{IndividualImpacts}
\end{figure}
}

{
Having constructed an individual model for cartel participation, we then observe its collective effects. Here, we consider how $C_1$ attempts to recruit members, and how people exposed to recruitment decide whether to join a cartel or not, as per Equation 5 of the manuscript. The goal here is to determine the level of heterogeneity in the recruitment process outcomes. For each value of the parameters, we simulate ten years of $C_1$ recruitment. Each week, a fraction of the $C_1$ members attempt to recruit new members for the cartel, exposing each ``recruiter'' to a single vulnerable person out of a large pool of potential recruits. Since the number of individuals that a cartel may recruit is too large (for instance, Mexico has approximately 17 million males between 13 and 30 years of age), we ignore finite size effects, competition with other groups in terms of recruitment and other elements such as physical proximity to a cartel. For each simulation, we vary the value of the social programs and quantify the number of people exposed to recruitment, as well as those who decide to join or not join a cartel (Figure \ref{IndividualImpacts}). 
}

{
However, this number drops to half with higher values of a social program (Figure \ref{IndividualImpacts}). Within 10 years, approximately 40,000 people would be exposed to $C_1$ recruitment and would choose not to participate for a sufficiently large value of a social program, compared to less than 18,000 when there are no social programs.
}

{
To simplify the cartel participation model (so that we can input some expression within the cartel dynamics and the control expression), the complexities and non-linearities of the cartel and the civil offer are captured with the function
\begin{equation}
\underbrace{g(v(t), C_i (t))}_\text{recruitment} = \rho b \E^{-\sigma v} C_i,
\end{equation}
capturing how cartel recruitment is reduced when a person receives $\text{US\$ } v$, depending on the \textit{efficiency} of social programs $\sigma$, the recruitment rate $\rho$ and a baseline rate $b$.
}

{
To react to a drop in recruitment, cartels could increase the share of forced recruitment (with higher values of $\pi_f$ and lower values of $\pi_v$ and $\pi_i$). Although this is an option for cartels, we speculate that the outcome for the group varies considerably. It is not the same loyalty to the group if its members joined because they perceived their gains to be larger inside the cartel than if they had joined against their will. Thus, we assume that forced recruitment happens with a much lower frequency than voluntary recruitment. 
}

{
Considering the large size of cartels in Mexico, many individuals will be exposed to cartel recruitment in the upcoming years. Thus, a critical difference is observed between the scenario in which many of those exposed to recruitment decide to join the group, and the scenario in which the civil offer outweighs the cartel offer. We observe here that with higher values of a social program, a higher income from legal activities, and by increasing the perceived risks of being killed or incapacitated, cartels won't be able to replace their losses simply by recruiting new members.
}

\subsection*{E - Budget and the reduction of violence}

{
Given the negative projections regarding increasing violence and harm caused by cartels in the coming years, we question whether investing in social and security programs is cost-effective. First, we analyse the scenario where the security budget is dropped. The government could ``save'' $\text{US\$ }1.11$ billion each year (currently used against $C_1$) and the same amount against $C_2$. However, in turn, $C_1$ and $C_2$ would become nearly 25\% larger. Without security programs, the number of homicides between $C_1$ and $C_2$ would increase by almost 90\% within the next decade. Although the costs of insecurity would slightly decrease in the short run (due to the money saved on security), the additional homicides and the extra cartel harm would surpass the savings within a couple of years. Against the current costs, we estimate that the 30-year impact of dropping security programs is over $\text{US\$ }74$ billion.
}

{
Similarly, we analyse the scenario without social programs. The government could ``save'' \text{US\$ }8.8 billion each year by dropping social programs. However, we estimate that social programs reduce the recruitment rate by 25\%. Without them, cartels would grow by nearly 50\% within a few years, leading to an increase in cartel-related harm and doubling the number of homicides over the same period. While reducing social programs might save money in the short term, the costs would surpass the savings within eight years, leading to a 30-year total cost exceeding \text{US\$ }53 billion above current levels. Security and social programs are vital for avoiding further deterioration in terms of violence in the country. Unfortunately, the impact is not sufficient to reduce the current cartel homicides or their harm and to reverse their growth trends. In the extreme scenario, if both social and security programs are dropped, a considerable amount of money would be saved, but the costs of the additional homicides and extra cartel harm would be higher in less than five years and, in the long run, would be the most expensive strategy. The 30-year cost of not investing in security and social programs would be 16\% higher than the current costs. 
}

{
We then consider the scenario of allocating the budget between security and social programs. Maintaining the same budget for social and security programs, we obtain a combined budget of $T = \text{US\$ }11.1$ billion each year. To consider a fixed budget strategy (meaning it does not change over time), we assume that a fraction $q$ of the budget is allocated to social programs, and a share $1-q$ is allocated against $C_1$ and $C_2$. Furthermore, we assume that the budget allocated to both cartels is the same. So we express the social budget as $qT$, which results in a weekly value of a social program of $v = qT / P$, and we express the security budget as $(1-q)T$, which results in a weekly budget of $u_i = (1-q)T/2$. We then consider values of $q \in [0, 1]$ to quantify the impact of the budget allocation on homicides and cartel harm (Figure \ref{ResultsFigureBudgetAllocation}). Results show that with most values of $q$, the number of homicides and cartel harm is higher than with the current budget distribution (observed at roughly $q = 80\%$).

\begin{figure}[!htbp]
\centering
\includegraphics[width=.95\textwidth]{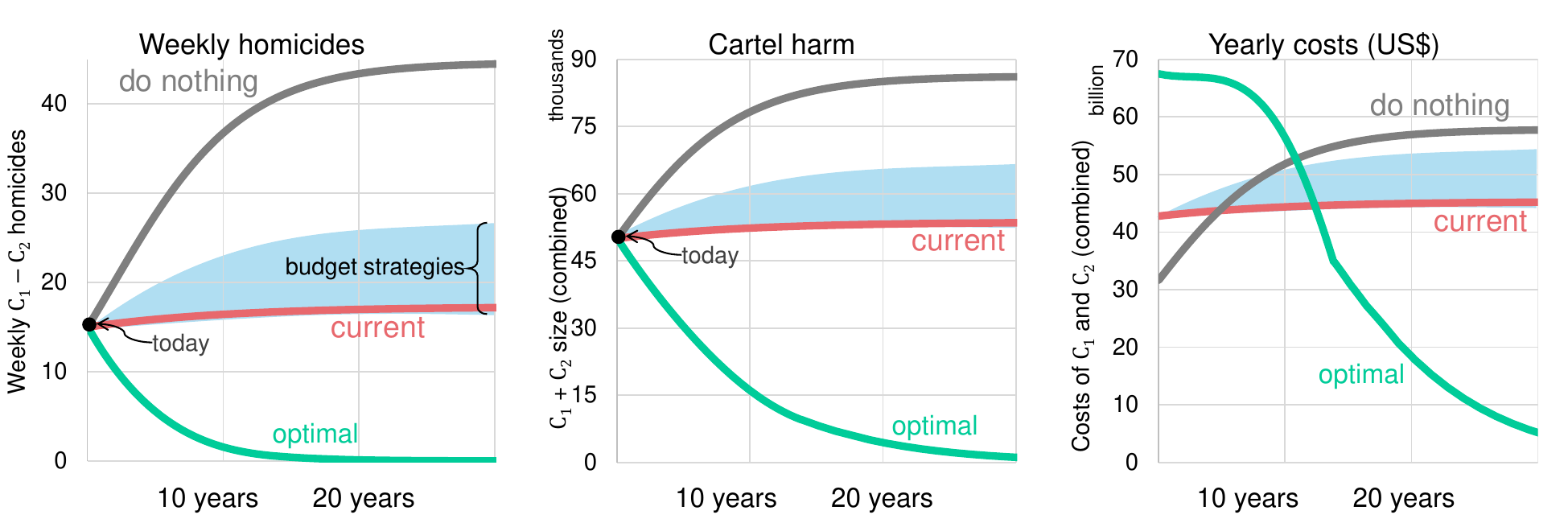}
\caption{Weekly number of homicides between $C_1$ and $C_2$ in the next three decades, depending on the budget used for social programs and the budget used in security programs (left). The blue area represents possible scenarios that could be observed by assigning different budgets to social programs and security programs. Cartel harm in the next three decades (centre). Total costs are considered, including social programs, security programs, the cost of homicides, and the cost of cartel harm (right).}\label{ResultsFigureBudgetAllocation}
\end{figure}

Although with a different budget allocation, there could be some decrease in terms of the number of homicides or cartel harm, the drop is minimal compared to the size of the problem.
}

\subsection*{F - Sensitivity analysis}

{
Our model depends on a set of parameters and assumptions. To analyse the relevance of the parameters' chosen values, we examine three budget scenarios: the current one, one with a zero budget, and the optimal budget strategy. For each scenario, we analyse the number of cartel homicides, cartel harm, and total financial costs within the next 30 years. 
}

{
\textbf{Lethality of cartels}. To analyse how sensitive the results are to the lethality of cartels, we vary the weekly number of homicides by $\pm 20\%$. Results obtained when varying the lethality of cartels show that although the baseline values change, the programs are equally ineffective in reducing violence or cartel harm, regardless of the lethality of the top two cartels (Figure \ref{SensitivityLethality}). 

\begin{figure}[!htbp]
\centering
\includegraphics[width=0.85\textwidth]{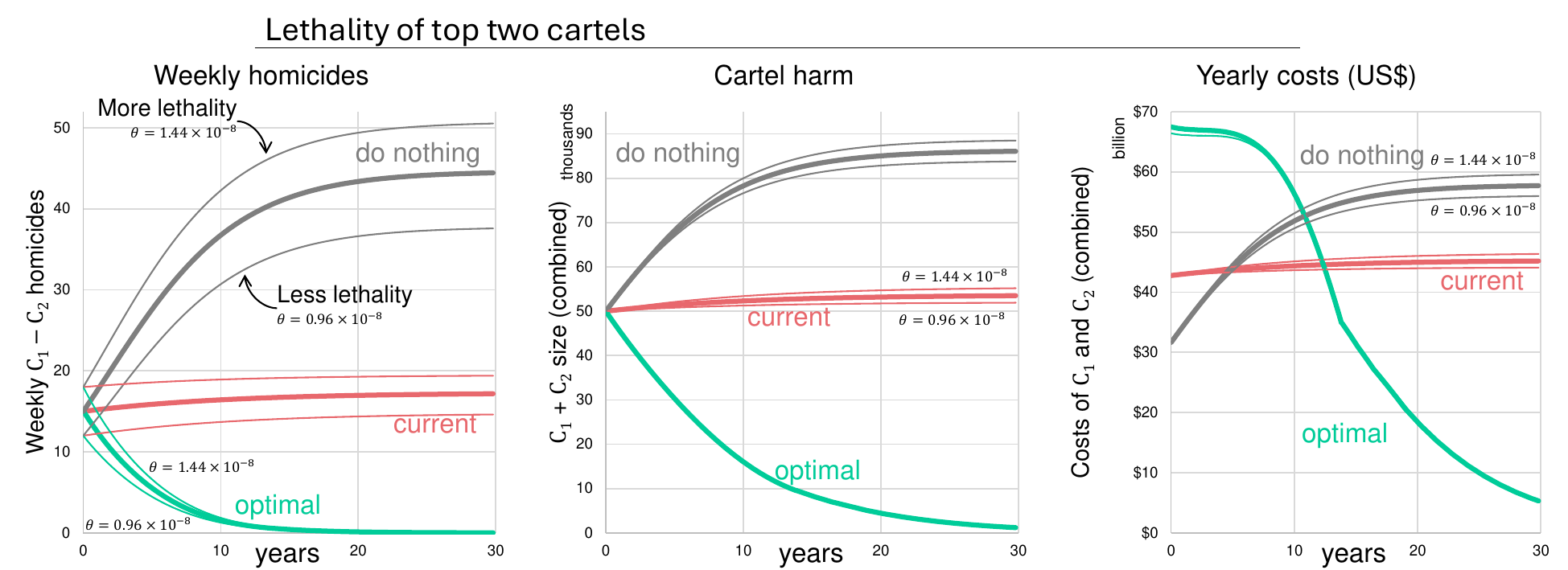}
\caption{Weekly number of homicides, cartel harm and costs depending on the budget invested in social and security programs (by colour). The thin lines correspond to different values of cartel lethality.  }\label{SensitivityLethality}
\end{figure}
}

{
\textbf{Asymmetric cartel targeting}. We study the effects of a strategy characterised by the possibility of asymmetrically targeting one of the two cartels instead of distributing the resources equally among the two. Specifically, once a cartel has been contained, resources can be subsequently allocated towards the next cartel. Ideally, this gives the state the advantage of having more resources and a localised strategy. We test this strategy by examining all combinations between a symmetric strategy (where $u_1 = u_2$) and the focused strategy (where $u_2 = 0$, but $u_1$ is doubled). Results show that a strategy that is focused more on a specific cartel yields a higher number of homicides for the first 20 years, but cartel harm remains much higher. By reducing the security budget used against any cartel, they grow faster, are engaged in more homicides and cause more harm. A similar scenario is observed when, instead of dealing with two large cartels (both of similar size), the country has to deal with a much larger cartel. Then, we consider whether assigning the budget unevenly against the top two cartels results in a faster reduction of homicides caused by cartel harm. We assign a share $x$ of the security budget $\psi$ against $C_1$, so $u_1 = x \psi$ and a share $1-x$ against $C_2$, so $u_2 = (1-x)\psi$. Then, we vary the values of $x \in [0, 1/2]$, which covers both cases due to the symmetry of the initial cartel size. Results show that there is no short-term advantage to fighting a single cartel. Fighting against a single cartel permits the rapid growth of another, and that effect takes a long time to stop. If the budget against cartels is not symmetric, it takes more than ten years to observe fewer homicides than the model forecasts under a symmetric budget. 

\begin{figure}[!htbp]
\centering
\includegraphics[width=0.75\textwidth]{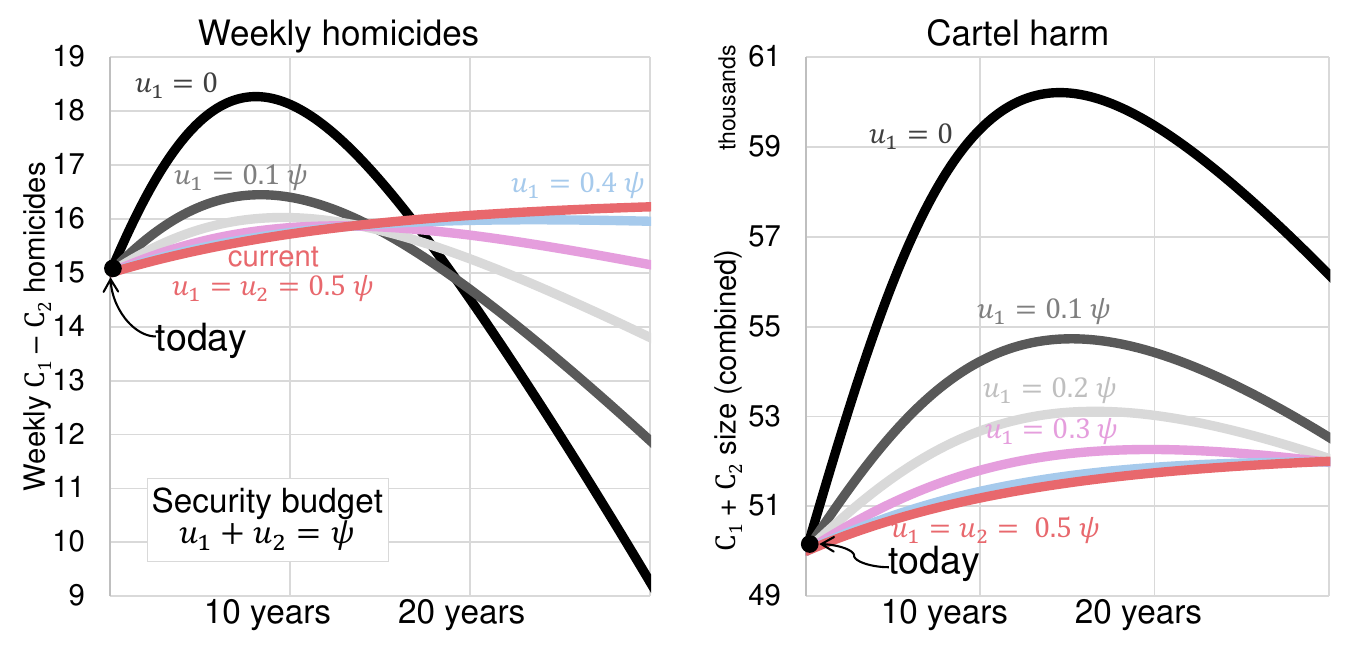}
\caption{Weekly number of homicides depending on the fraction of the security budget that is used against $C_1$ (left) and cartel size (right).  }\label{Results30YearsAssymetricStrategy}
\end{figure}
}

{
\textbf{Asymmetric cartel size and targeting.} Finally, to analyse how sensitive the results are in terms of having asymmetric cartels, we analyse the base case in which $C_1(0) = c_2(0) = 25,000$ and an asymmetric case, in which $C_1(0) = 31,500$ and where $C_2(0) = 18,500$. In both the asymmetric and symmetric cases, $C_1(0) + C_2(0)$ remains fixed, but we alter their initial size by more than 25\%. Results for the asymmetric case show minor variations compared to the symmetric case (Figure \ref{SensitivityAssymetry}). Thus, even if the initial size of cartels is the same or if one is 70\% larger than the other, we still have an increasing level of murders and cartel harm with the current social programs, a fast increase with no programs and a rapid decrease with an unconstrained budget. 

\begin{figure}[!htbp]
\centering
\includegraphics[width=0.85\textwidth]{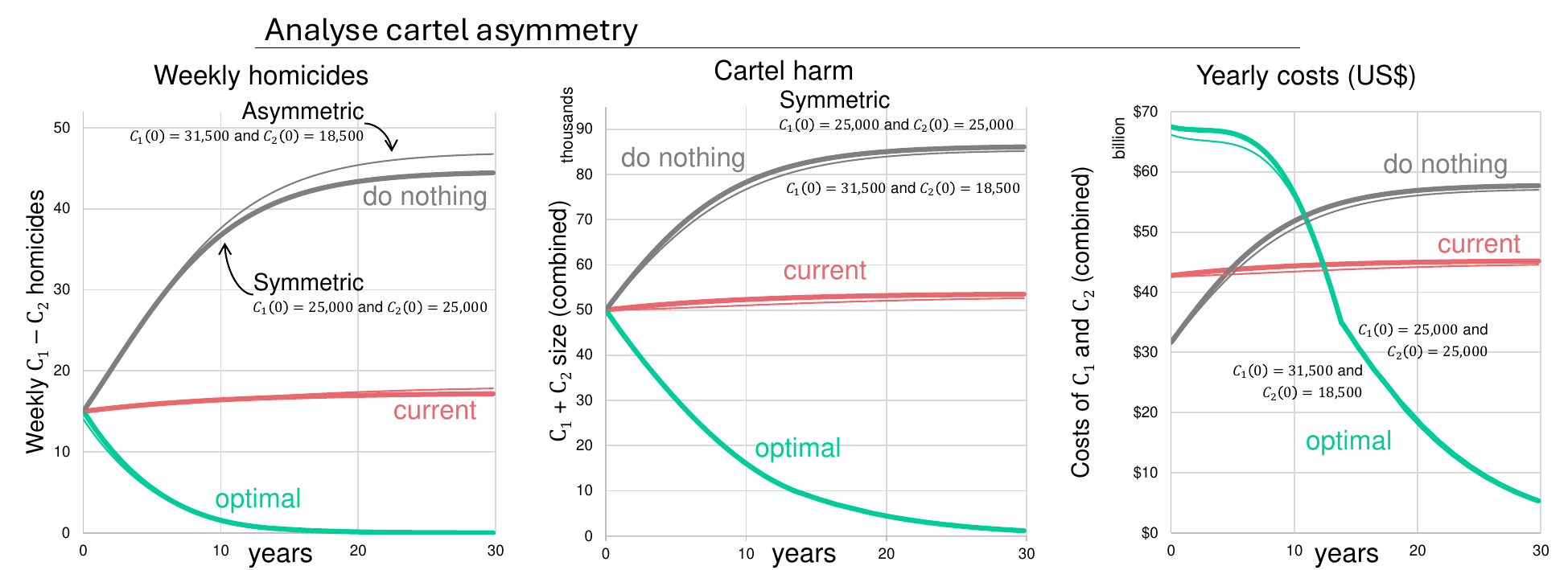}
\caption{Weekly number of homicides, cartel harm and costs depending on the budget invested in social and security programs (by colour). The thin lines correspond to different values of cartel lethality.  }\label{SensitivityAssymetry}
\end{figure}
}

\subsection*{G - Cartel costs, revenue and profit}

{
Cartels are dynamic organisations, capable of adjusting their strategy in response to changing conditions. For example, the state could design social programs to prevent cartel recruitment, invest in security programs or conduct joint international operations, increasing the members' risk of incarceration. However, cartels react to these changes. They may modify recruitment strategies or even shift their territorial presence in response to government pressures. These responses correspond to the cartels' resilience and may reduce the effectiveness of policy interventions. For example, a cartel might react to increasing values of social programs by adjusting wages offered as their premium, which could reduce their effectiveness. Observing the activities of organised crime from the perspective of an enterprise, we model the cartel's revenue and costs and determine how they should react to increased social programs or enhanced security.  
}

{
To capture how a cartel reacts to different social or security programs, we consider how its profit changes. According to the cartel participation model, a person compares the cartel offer with the civil offer (as in Equation 5 of the manuscript) by computing the respective monetary gains and associated risks by combining
\begin{eqnarray}
\Sigma  &=& \underbrace{\alpha}_{\text{baseline}}  +\underbrace{ \gamma r -  \delta \kappa_P - \eta \kappa_K }_{\text{cartel offer}} - \underbrace{ ( \gamma w  + \gamma v )}_{\text{civil offer}} \\
&=& \alpha + \gamma (r - w - v) -  \delta \kappa_P - \eta \kappa_K.
\end{eqnarray}
To maintain recruitment, a cartel must balance its offer and the civil offer, keeping the values of $\Sigma$ unchanged, so they are forced to increase their premium. Thus, the impact of social programs (and of enhanced security) is that recruitment becomes more challenging. For example, if social programs increased by one dollar, the cartel would need to match that expenditure to maintain the recruitment. A similar phenomenon occurs when security programs increase, leading to a higher risk of incapacitation. 
}

{
When a cartel has to pay a higher premium to recruit and maintain its members, it could react by increasing its total income. For example, by increasing the cost of some illicit drugs and transferring part of the extra fees to consumers \cite{gallet2014can}. However, cartels face intense competition from rival Mexican groups and transnational organisations that could potentially take their place in the drug trade. Thus, despite its high profitability and inelastic demand \cite{gallet2014can}, cartels are not in a position to unilaterally raise drug prices in response to domestic policies \cite{stiglitz1979equilibrium}. Additionally, cartels are active in many activities beyond taking part in the illicit drug supply chain, including extortion \cite{santos_cid2025rehenes}, kidnapping, fuel theft (commonly referred to as ``huachicol''), and involvement in markets such as lime and avocado production. Parts of the cartel's income depend on the legal market, for example, in the trade of petrol, for which it is not feasible for the cartel to increase the price, or depend on the capacity to pay, as in the case of kidnapping and extortion. For different parts of their activities, increasing income is not a matter of just a few simple steps; otherwise, they would have already taken those steps. Therefore, we assume that a cartel cannot simply increase its profit to compensate for changes in social or security programs. Instead, their optimal reaction involves adjustments to their total costs. 
}

{
We consider first the revenue of $C_1$ generated by all its members and through all activities, including fuel trade, extortion, or the trade of illicit drugs. Regarding the revenue, we model it as a function of the number of members, $N$. Due to coordination costs, management capacity and a limited market size (for illicit drugs) and a finite number of potential victims (for example, companies to victimise, people to kidnap, petrol to steal, etc), the cartel has diminishing marginal returns to labour \cite{shephard1974law}. One expression to capture the diminishing returns of labour is  
\begin{equation}
Y(N) = \alpha N^\beta,
\end{equation}
where $Y$ is the yearly income in US\$ for that cartel, and $\alpha>0$ and $0<\beta < 1$ are two model parameters. This expression is also a Cobb–Douglas function taking into account only labour. Considering a large sample across industries, their revenue was approximated by a sublinear expression regarding the number of employees \cite{west2018scale}. Across industries in the US and China, values of $\beta$ between 2/3 and 3/4 were observed \cite{zhang2021scaling}. The value of $\beta$ for organised crime is not a parameter that can be derived from data, so we take a range of possible values of $\beta \in  [2/3, 3/4]$.
}

{
We then consider the total costs for $C_1$ as a function that depends on the number of members. We decompose the costs into a combination of wage $r$ and other costs $\omega$ per person (including, for example, uniforms, weapons, vehicles, and other operational expenses), and we assume that these costs increase linearly with the number of members. Thus, the average cost per employee can be expressed as
\begin{equation}
r + \omega,
\end{equation}
so the costs for the cartel $X(N)$ can be expressed as
\begin{equation}
X(N) = (r + \omega) N.
\end{equation}
}

{
Subtracting the costs from the revenue, we obtain that the profit of $C_1$ is given by
\begin{equation}
R(N) =  \alpha N^\beta - (r + \omega)N.
\end{equation}
The cartel $C_1$ follows a strategy that aims to maximise its profit, so its objective is to pick the number of members $N$ such that $R(N)$ is maximum. Its optimal number of members is achieved when the marginal returns to income equal the marginal costs \cite{stiglitz1979equilibrium}. With this principle, the optimal size is obtained when
\begin{equation}
\alpha \beta N^{\beta-1} = r + \omega,
\end{equation}
from which we obtain that
\begin{equation}
\alpha^\star = \frac{r + \omega}{\beta} N^{1-\beta}.
\end{equation}
We use that value of $\alpha^\star$ to capture the income structure of $C_1$ and assume it is fixed, despite changes in social or security programs. 
}

{
Based on the revenue and costs structure of the cartel, we can estimate its optimal size considering the effect of social or security programs. For social programs, for example, when a person receives $v$ dollars, the cartel matches that offer, and their optimal strategy is 
\begin{equation}
\alpha^{\star} \beta N^{\beta-1} = r + \omega+v,
\end{equation}
from which we get that the optimal number of cartel members for $C_1$, as a function of the value of social programs, is given by
\begin{equation} \label{DimSize}
N^{\star} = \left( \frac{r + \omega+v}{ \beta \alpha^{\star}} \right)^{1/(\beta-1)}.
\end{equation}
}

{
The expression \ref{DimSize} captures the optimal number of members of $C_1$ varying the value of social programs. For higher values of a social program, $v$, the costs incurred by the cartel to recruit and maintain members increase. In turn, the optimal group size drops considerably. For values of $\beta \in  [2/3, 3/4]$ interval, the optimal size $N^\star$ ranges from 
\begin{equation}
N^{\star}(v) \in \left[ \left( \frac{3 \alpha^{\star}}{4(r + \omega + v)} \right)^4, \left( \frac{2 \alpha^{\star}}{3(r + \omega + v)} \right)^3 \right].
\end{equation}
Thus, the best strategy for a cartel is to reduce its size in response to any increase in social programs. 
}

{
To examine the impact of social programs on cartel size, we utilise the estimates observed for cartels in Mexico. We take $N^\star = 25,000$, the number for $C_1$ members \cite{PrietoCampedelliHope2023} and assume that $C_1$ has maximised their profit with that size. We take $\omega$ to be 20\% of the value of $r$, so the costs of a cartel member are $5/6$ the salary and $1/6$ the other expenses (Figure \ref{CartelStrategy}). We then estimate that the yearly revenue of the $C_1$ cartel is between US\$520 million (for a large $\beta = 3/4$) and US\$585 million (for a small $\beta = 2/3$). Within this range, the profit of $C_1$ is between US\$130 million and US\$195 million. We use these values as the ``current state of affairs'' and then vary social programs, assuming that the costs for the cartel to recruit members change. With $v = v_0 = \text{US\$ }832 $ per person, then the optimal size is $N^{\star} = 25,000$. 

\begin{figure}[!htbp]
\centering
\includegraphics[width=0.95\textwidth]{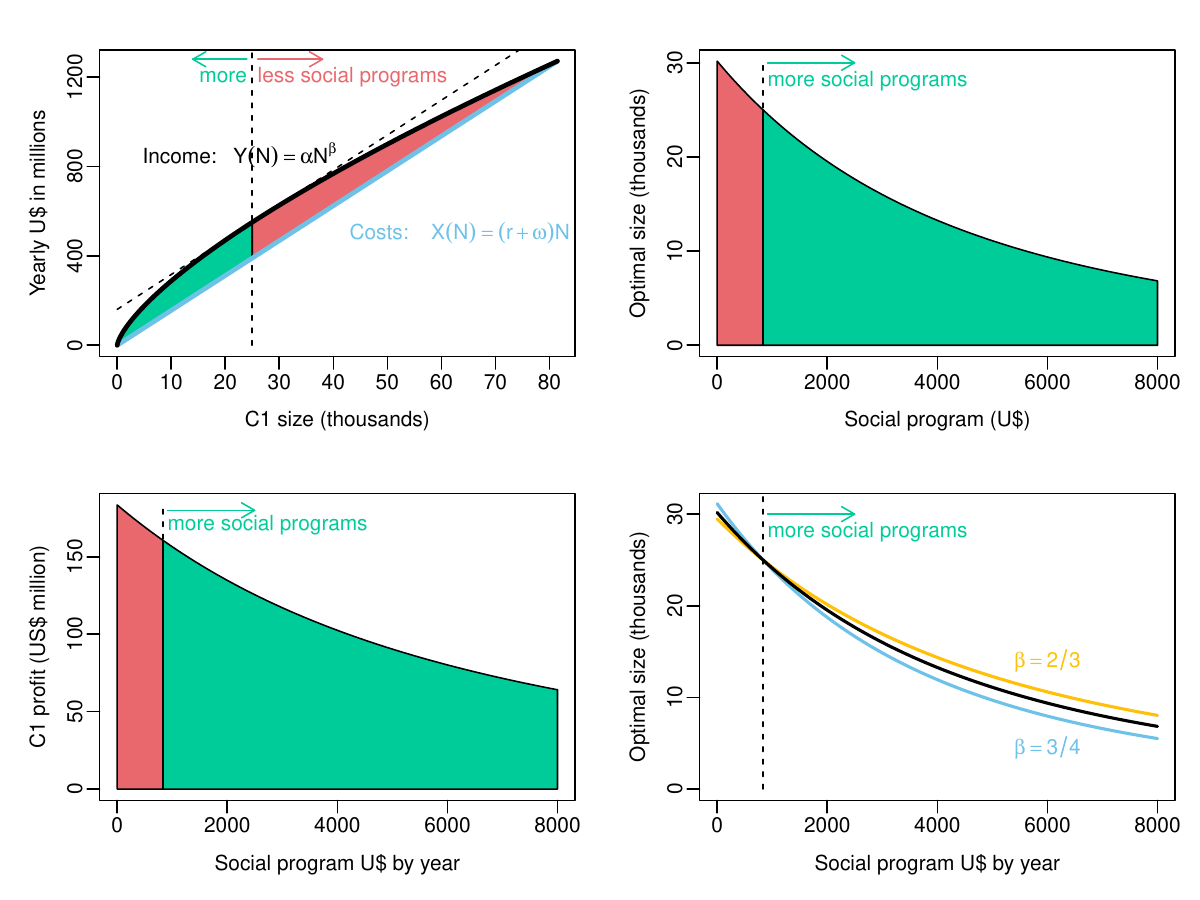}
\caption{Top left - Income (thick line) and costs for $C_1$ per year in US\$ (vertical) for different sizes of the cartel (horizontal). The dashed line corresponds to the estimated size. The red area corresponds to segments where the cartel has reasons to increase its number of members, as they have higher revenue (so marginally, each new member produces more than it costs). The tilted dashed line is drawn to identify the marginal costs at the optimum size. Top right - Optimal cartel size (vertical axis) as a function of the social programs (horizontal). The vertical dashed line is the current value of social programs. Bottom left - Profit of $C_1$ (vertical axis) depending on the value of social programs (horizontal axis). Bottom right - Optimal group size (vertical axis) for three values of the $\beta$ parameter, depending on social programs (horizontal). }\label{CartelStrategy}
\end{figure}
}

{
The optimal size for cartels decreases with the introduction of more social programs, which then has a triple effect. For example, if social programs double, then $v = 2  v_0$. In that case, the optimal size of cartels is $20,900$. Thus, if social programs double, recruiting and maintaining members becomes more expensive, so the optimal size is 16\% smaller. With a smaller $C_1$ (and more costly recruiting and keeping members), the profit for cartels is also smaller. If social programs double, we estimate that the profit for cartels is 13\% smaller (Figure \ref{CartelStrategy}). Additionally, if both cartels $C_1$ and $C_2$ have simultaneously a smaller strategic size (since both groups have a more difficult recruiting process), the term $\theta C_1 C_2$, which quantifies the weekly number of homicides between their members, drops by more than 30\%. Thus, by making recruiting more costly for cartels, they have a smaller optimal size, which directly drops their profit and eventually results in a drop in their homicides.   
}

{
On the other hand, without social programs (with $v = 0$), the optimal size of $C_1$ would be nearly 30,000 members, as recruiting and maintaining members would be cheaper, and their profit would be 13\% larger. In this scenario, the number of weekly homicides between $C_1$ and $C_2$ would increase by 44\%. Thus, for a cartel, higher social programs translate into higher personal costs for recruiting and maintaining its members. With higher costs, a cartel has a smaller strategic size and a reduction in its profit.
}

\subsection*{H - Cartel shocks in the short and the long run}

{
Some security policies, such as the beheading of criminal organisations, may have unintended consequences, increasing intra-cartel fighting \cite{calderon2015beheading}. Rival cartels often attempt to usurp territories and strategic highways after crackdowns have weakened a group \cite{dell2015trafficking}. Cartels adapt their areas and goals, so their conflict evolves \cite{TrejoWhyDidDrug2018}. Once two cartels have engaged in a conflict, it frequently follows retaliatory dynamics, escalating until one of the groups is defeated and displaced from a territory or fragmented into small cells \cite{calderon2021organized}. This retaliation dynamic creates patterns which might escalate quickly \cite{egesdal2010statistical, short2014gang}. One of those cases was the outbreak of violence in Sinaloa in 2024 following El Mayo Zambada's arrest, which resulted in some of the highest levels of violence in Mexico in recent years \cite{arocha2025sinaloa}. Only weeks after the arrest, the number of homicides in Sinaloa tripled \cite{rivera2024homicidio}. Therefore, it is crucial to comprehend the shocks that impact the system and assess their short- and long-term effects.
}

{
Here, we model the shocks associated with violence. They may be triggered by many factors (such as a cartel beheading, a big drug seizure or any major event) at some time $t_0$. It was observed that most of the effects following the removal of a leader from a cartel happen within the first six months \cite{calderon2015beheading}. We extend our model by capturing the changes in cartel size with the expression
\begin{eqnarray}\label{MasterEquation}
\dot{C}_i (t) & = & \underbrace{g(v(t), C_i (t))}_\text{recruitment} - \underbrace{f(u_i (t), C_i (t))}_\text{incapacitation} - \underbrace{\theta(t) C_i (t) C_j (t)}_\text{conflict} - \underbrace{\omega C_i^2 (t)}_\text{saturation}. 
\end{eqnarray}
The term $\theta(t) C_i C_j$ captures, with the (time-varying) lethality function $\theta(t)$, how the conflict between cartels $i$ and $j$ evolves. To model a shock, we express the lethality function as
\begin{equation}
\theta(t) = \theta_0(1 + \kappa I_{[t_0, t_0+d]}(t)), 
\end{equation}
where the binary function $I_{[t_0, t_0+d]}(t)=1$ if $t$ is inside the $[t_0, t_0+d]$ interval and zero elsewhere, where $d$ marks a six-month interval after some shock which occurred at time $t_0$, and $\kappa$ captures the magnitude of the shock. With $\kappa =2$, for example, the lethality of the $C_1$ - $C_2$ conflict triples for six months and then returns to its baseline value (Figure \ref{Retaliation}).

\begin{figure}[!htbp]
\centering
\includegraphics[width=0.95\textwidth]{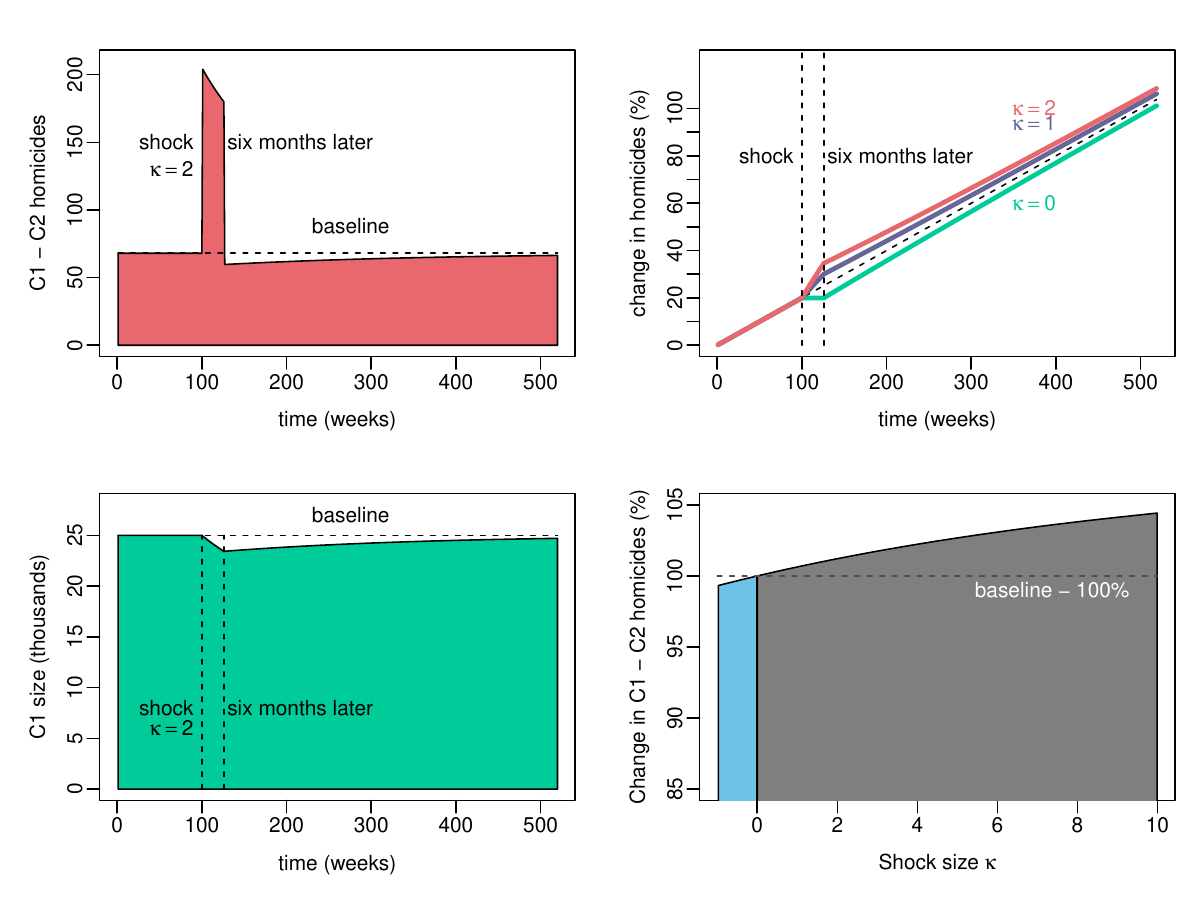}
\caption{Top left - Weekly number of $C_1$ - $C_2$ homicides (vertical) for ten years (horizontal). At week $t_0 = 100$, there is a shock, with intensity $\kappa = 2$, so the number of homicides triples (so a similar shock to the system to what was observed in Sinaloa in 2024). The shock lasts for six months, after which the value of $\theta$ returns to its baseline. However, since both cartels are now smaller (due to the additional homicides during the shock), it takes time for them to return to their baseline values. Top right - Cumulative number of homicides (vertical) compared to the baseline scenario without a shock. Bottom left - Size of the $C_1$ cartel (vertical) for ten years (horizontal). After the shock in week $t_0 = 100$, the cartel is slightly smaller than its pre-shock levels, but recovers through recruitment. Bottom right - Change (in \% in the vertical axis) between the number of casualties observed for 30 years against the scenario without a shock, depending on the intensity (horizontal). }\label{Retaliation}
\end{figure}
}

{
We assume that cartels were in a stable state before the shock, meaning they were able to compensate for their losses through recruitment. Starting with the shock at time $t_0$, there is a high number of homicides, which slowly drops, even during the high-intensity moments during the shock period, resulting from the decrease in the size of both cartels. After six months, when the values of $\theta(t)$ return to their original value, the number of casualties drops below the baseline, as both cartels now have fewer members and are capable of creating a smaller impact on their enemies than before the shock. Yet, what we observe is that precisely since the number of homicides is smaller than the baseline, cartels are capable of replacing their losses during the shock by recruiting more members. Within a couple of years after the shock, both cartels have roughly the same size and thus, create a similar number of homicides as before the shock. 
}

{
The magnitude of the shock, $\kappa$ is critical, particularly for the short-term report of the number of crimes. For example, with $\kappa = 2$, which means that the number of $C_1$ - $C_2$ homicides triples (similar to the Sinaloa shock in 2024), then during the year of the shock, the reported homicides increase by more than 90\%. However, since cartels recover after a shock by recruiting new members, the year after the shock, there is a decrease of 3\% in the number of homicides with respect to the baseline (so fewer homicides than before the shock). Combining the costs, the long-term impact of that shock is an increase of only 1.2\% in the 30-year (cumulative) number of homicides. The number of members of $C_1$ and $C_2$ is so large that shocks of a reasonable magnitude (where reasonable here means even twenty or thirty times more violence for six months) are far from reducing their size beyond a critical size (meaning, beyond a size from which they cannot recover). 
}

{
A shock to the conflict between two cartels could also lead to a temporary peace between them. In the scenario, expressed with $\kappa = -1$, where the shock also lasts for six months and returns to its baseline, the number of homicides for that year drops 46\%. However, in the long run, the number of $C_1$ - $C_2$ homicides drops less than 1\% in that scenario.
}

{
There are thus two ways to observe the shock between two cartels. In the short term, a shock may rapidly increase the number of homicides they create. The retaliatory dynamics between cartels produce alarming waves of violence, which demand rapid responses and a quick operation to de-escalate the conflict. Yet, in the long run, most of the homicides between cartels are not observed during those high-intensity moments, but are the result of a long and sustained fight, which creates a sustained number of homicides for years.
}

{
Here, we have modelled changes in cartel size related to recruitment, incapacitation, conflict and saturation \cite{PrietoCampedelliHope2023}. That means that by expressing changes in cartel size by $\dot{C}_1$ as a combination of elements, then we can detect if a size is a fixed point of the system (so one where $\dot{C}_1 = 0$ and if it is stable (where small perturbations to the size eventually return to the same equilibrium size. What we observe here with the shocks is that for a cartel at its equilibrium size, these shocks are perturbations that are too small actually to alter their equilibrium size.
}

{
\paragraph{Are cartel shocks desirable?} We formulated the government's objective by adding the financial burden of the violence that cartels create, the impact of their presence and the combined costs of security and social programs. We added these costs for a long period to understand the long-term costs of different policies. Thus, we express the objectives as 
\begin{eqnarray} \label{ObjectiveEqRep}
Q&:=& \int_0^\infty \E^{-rt} \left( \underbrace{2 s \theta(t) C_1(t) C_2(t)}_{\text{homicides}} + \underbrace{ h ( C_1(t) + C_2(t))}_{\text{cartel harm}}+ \underbrace{ u_1(t) + u_2(t)}_{\text{security}} + \underbrace{P v(t)}_{\text{social}} \right) \,\Dt.
\end{eqnarray}
Shocks create two simultaneous effects. On the one side, they increase the number of homicides, which means that the first term in the integral of equation \ref{ObjectiveEqRep} is larger. However, they decrease the size of cartels, so the second term, the cartel harm, is smaller. The tradeoff between cartel homicides and harm is precisely the motivation to add both terms to the long-term costs.  
}

{
The results of the model show that the long-run costs of a cartel shock, within a reasonable size (i.e., $\kappa \in [-1, 10]$), result in a roughly equal burden on society. With peace that only lasts for six months, cartels get larger, but there are six months with a drop in homicides, increasing the long-run costs by 0.05\%. So, an artificial peace agreement between cartels does not reduce their costs. On the other hand, with a shock of $\kappa = 2$, violence triples, causing cartels to shrink slightly (by only 1.5\%, which is not a significant impact on the group), thereby temporarily reducing their harm. However, the costs of homicides compensate for that drop, increasing by over 4\%. With a shock of $\kappa = 2$, the long-run costs are 0.09\% smaller. Thus, cartel shocks (both positive and negative) do not reduce the long-run costs and the harm that they impose on the country.
}

\subsection*{I - Cartel fragmentation}

{
A consequence of some security policies is that they may alter the links between cartels \cite{atuesta2018fragmentation}. The arrest of El Mayo Zambada in July 2024, for example, triggered a violent rupture within the two factions of the Sinaloa Cartel: Los Chapitos and La Mayiza \cite{arocha2025sinaloa}. The links between cartels and their factions are highly complex \cite{PrietoCampedelliHope2023} and evolve in response to shifting policies \cite{Esberg_2025}. These groups frequently form and dissolve alliances, reshaping the broader cartel landscape. For instance, in 2010, the Gulf Cartel temporarily allied with the Sinaloa Cartel to counter the growing threat posed by Los Zetas, an alliance that contributed to a surge in brutal violence across Mexico \cite{jones2022mexico}. Here, we analyse the case of cartel fragmentation and quantify its impacts in terms of cartel harm and homicides.
}

{
Cartel fragmentation may be observed as a permanent shock, rather than one that lasts only for a short period. Those factions that split and start a fight will rarely reunite as a unified cartel. Cartel fragmentation can be modelled by changes in the function $\theta(t)$ for the interaction between faction $A$ and faction $B$ of some cartel, $C_1$. Taking the two factions as separate cartels, we can model their dynamics, with their lethality corresponding to
\begin{equation}
\theta(t) = \theta_0 I_{[t>t_0]}(t) , 
\end{equation}
where the binary function $I_{[t>t_0]}(t)=1$ after time $t_0$ which is when the cartel has fragmented (and the function is zero before $t_0$, when the factions were united), and for some value of $\theta_0$ which captures the lethality between the two factions $C_A$ and $C_B$. We assume the lethality $\theta_0$ to be of similar magnitude to the fight between the top two cartels, $C_1$ and $ C_2$. That is, the fight between two factions of a fragmented cartel is assumed to be as violent as if they were two conflicting cartels.
}

{
To quantify the impact of cartel fragmentation, we consider a group composed of two factions, $A$ and $B$, of similar size, which are in equilibrium. At some point, $t_0 = 100$ weeks, they fragment into two rival cartels. We compare the scenario without cartel fragmentation as the baseline to quantify the impact in terms of life lost and cartel harm. The short-term implications of cartel fragmentation are a rapid escalation in homicides. The retaliatory dynamics between the factions produce alarming waves of violence, particularly since previous allies may have more information on how to inflict greater damage on their (now) enemies and how to make it more visible and strategic. 
}

{
In the long run, both factions will have a smaller equilibrium size (since now they have to combat another organisation), but the country will experience a higher number of homicides. Results of the model show that with fragmentation, both factions will have a smaller equilibrium size, which has approximately 14\% fewer members. Thus, the cartel harm is reduced since the number of cartel members is smaller. However, due to fragmentation, there is a sharp increase in cartel homicides, which almost compensate for the drop in cartel harm. The long-run costs of cartels, combining their harm and homicides, drop by approximately 1\% with fragmentation. 
}

{
In the context of cartel fragmentation, violence levels can escalate significantly, as evidenced in Sinaloa \cite{rivera2024homicidio}. Moreover, the trajectory of conflict between emerging factions remains uncertain. Some cartels, such as the CJNG, are known to incorporate ``orphan cells'' from splintered organisations into their structure \cite{jones2022mexico}. For instance, while the Sinaloa Cartel has experienced intense clashes with the CJNG in recent years, its internal division into Los Chapitos and La Mayiza opens the possibility that one of these groups could eventually align with CJNG \cite{dea2025ndta}.
}

{
Our model for cartel fragmentation is a simplified description of a much more complex process. However, it highlights two crucial elements to observe cartels quantitatively. First, some cartels are enormous organisations with thousands of members. Because of fragmentation, each faction will lose approximately 2\% of its members each year due to the fight with the other faction. And second, the burden of cartels is a combination of the homicides they inflict upon other cartels, and all the harm they have on society, including extortion, kidnapping, the trade of petrol, illicit drugs and others. With cartel fragmentation, there is a trade-off between reducing cartel size and increasing cartel homicides. Thus, it is critical to consider adequate indicators to quantify the harm of cartels in the country, including not only the number of homicides, but also the number of missing people, kidnappings, as well as the number of companies and people suffering extortion and other types of crime.
}

\subsection*{J - Optimality conditions}

{
The problem we are dealing with is an optimal control problem with an infinite time horizon. It can be solved using the standard Maximum Principle (see~\cite{grass_optimal_2008}), which provides a set of necessary optimality conditions for optimality. These are formulated in the following. The Lagrangian is given by
\begin{eqnarray}
\mathcal{L} & = & w_1 s C_1 C_2 + w_2 h ( C_1 + C_2)+ u_1 + u_2  + P v \nonumber \\
&& + \lambda_{1} \left( \rho e^{-\sigma v} C_1 - \eta (u_1 C_1)^\pi - \theta C_1 C_2 - \omega C_1^2 \right) \nonumber \\
&& + \lambda_{2} \left( \rho e^{-\sigma v} C_2 - \eta (u_2 C_2)^\pi - \theta C_1 C_2 - \omega C_2^2 \right) \nonumber \\
&& + \mu \left( Pv+u_{1}+u_{2}-M \right) ,
\end{eqnarray}
where $\lambda_{i}$ denotes the adjoint variables (dynamic shadow prices: increase of the objective function if $C_i$ increases by one marginal unit) of $C_{i}$, and $\mu$ represents the Lagrangian multiplier for the budget restrictions.
}

{
By derivative of $\mathcal{L}$ with respect to the control variables, the first-order conditions are
\begin{eqnarray}
\frac{\partial \mathcal{H}}{\partial u_{1}} & = & 1 - \lambda_{1} \pi \eta (u_1 C_1)^{\pi -1} C_{1} + \mu =0 \qquad \Longrightarrow \qquad u_1 = \frac{1}{C_{1}} \left( \frac{1 + \mu}{C_{1}\lambda_{1} \pi \eta} \right)^{\frac{1}{\pi -1}} \nonumber \\
\frac{\partial \mathcal{H}}{\partial u_{2}} & = & 1 - \lambda_{2} \pi \eta (u_2 C_2)^{\pi -1} C_{2} + \mu =0 \qquad \Longrightarrow \qquad u_2 = \frac{1}{C_{2}} \left( \frac{1 + \mu}{C_{2}\lambda_{2} \pi \eta} \right)^{\frac{1}{\pi -1}} \nonumber \\
\frac{\partial \mathcal{H}}{\partial v} & = & P - \sigma \rho e^{-\sigma v} \left( \lambda_{1} C_1 + \lambda_{2} C_2 \right) + P \mu =0 \qquad \Longrightarrow \qquad v =- \frac{P}{\sigma} \ln \left( \frac{1 + \mu}{\sigma \rho \left( \lambda_{1} C_1 + \lambda_{2} C_2 \right)} \right).
\end{eqnarray}
From the first two equations we find the relation between $u_{1}$ and $u_{2}$ in the optimum, which is
\begin{eqnarray}
\frac{u_1}{u_2} & = & \left( \frac{\lambda_{2}}{\lambda_{1}} \right)^{\frac{1}{\pi -1}} \left( \frac{C_{2}}{C_{1}} \right)^{1- \frac{1}{\pi -1}} .
\end{eqnarray}
This means that the spendings for the two cartels are divided amongst them by the dynamic marginal rate of substitution (represented by the term $\frac{\lambda_{2}}{\lambda_{1}}$) normalised by the relative cartel sizes.
}

{
The adjoint variables are defined by the adjoint equations
\begin{eqnarray}
\dot{\lambda}_{1} & = & \left( r - \rho e^{-\sigma v} + \eta \pi (u_1 C_1)^{\pi -1} u_{1} + \theta C_2 +2 \omega C_1 \right) \lambda_{1} - w_1 s C_2 - w_2 h+ \lambda_{2} \theta C_2 \nonumber \\
\dot{\lambda}_{2} & = & \left( r - \rho e^{-\sigma v} + \eta \pi (u_2 C_2)^{\pi -1} u_{2} + \theta C_1 +2 \omega C_2 \right) \lambda_{2} - w_1 s C_1 - w_2 h+ \lambda_{1} \theta C_1
\end{eqnarray}
and limiting transversality conditions. The integration of both cartels in the dynamic solution works on two levels: (i) Term $-w_1 s C_j$: The homicides term in the objective function is multiplicative in the size of both cartels. Therefore, the effect of a marginal unit of $C_{i}$ on the homicides has to be multiplied by $C_j$, i.e., the effect of one cartel member is proportional to the other cartel size. (ii) Term $+\lambda_j \theta C_j$: Homicides, on the other hand, also reduce the size of both cartels, which has to be evaluated positively. Thus, in the case of reducing the size of one cartel, the police profit from the size of the other cartel. This effect increases the value of one marginal unit of the cartel.
}

{
Finally, the standard complementary slackness condition guarantees that the budget constraint holds on the entire solution path:
\begin{eqnarray}
0 & = & \mu \left( Pv+u_{1}+u_{2}-M \right).
\end{eqnarray}
Details on the numerical solution procedure for this class of problems can be found in \cite{grass_optimal_2008}. 
}

\subsection*{K - More efficient social programs}

{
We modelled the effect of social programs regarding their cost and how they can prevent cartel recruitment. Strategic investments in education, job training, and community development could weaken the appeal of cartels, fostering long-term social resilience. However, social programs could become more effective in preventing cartel recruitment by focusing on the most vulnerable populations, such as impoverished youth, marginalised communities, and individuals with limited access to education or employment opportunities. By tailoring interventions to address specific risk factors like economic instability, lack of social support, and exposure to violence, these programs can offer viable alternatives to organised crime. Thus, we model here what could happen if social programs were more effective at preventing cartel recruitment. 
}

{
To analyse how sensitive the results are to the model for social programs, we compare the number of homicides, cartel harm and costs for 30 years under the current strategy, the scenario in which the government does nothing and the case with an unconstrained budget. We vary then the premium salary offered by cartels by $\pm 20\%$, so we consider the set in which $b_m = 1.2075$ and $\sigma_m = 2.27 \times 10^{-4}$ (where the social programs have a smaller impact) and the set with $b_M = 1.3321$ and $\sigma_M = 3.29 \times 10^{-4}$ (where social programs have a larger impact). Results show that, although there is an impact in the scenario where no budget is invested in social and security programs, it is still worthwhile to invest in them (Figure \ref{SensitivitySocialPrograms}). Also, in the long run, it is convenient to invest more resources in social and security programs, regardless of whether the efficiency of social programs to stop cartel recruitment varies $\pm 20\%$. 

\begin{figure}[!htbp]
\centering
\includegraphics[width=0.85\textwidth]{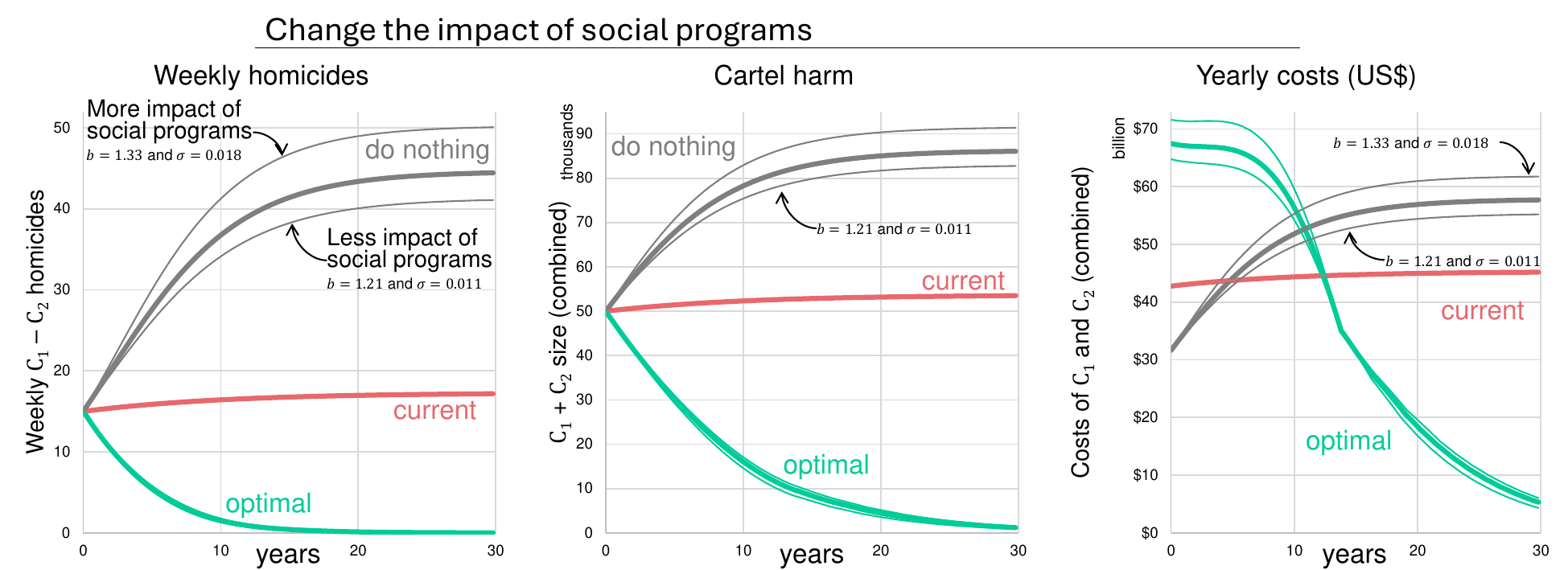}
\caption{Number of homicides, cartel harm and costs depending on the budget invested in social and security programs (by colour). The thin lines correspond to different values of the efficiency of social programs.  }\label{SensitivitySocialPrograms}
\end{figure}
}

{
However, the results of the analysis also show that with the current budget, making social programs more effective at preventing cartel recruitment has a negligible impact in terms of the number of homicides, cartel harm or the total costs. Thus, increasing the impact of social programs to prevent cartel recruitment has little effect, considering the current budget restrictions.
}

\subsection*{L - Impact of more efficient security programs}

{
We model the effect of security programs on their impact in incapacitating cartel members and their associated costs. Between 2019 and 2022, approximately 120,000 yearly incapacitations were recorded, with the majority related to minor offences such as theft, injuries, and domestic violence \cite{ENPOL}. Additionally, it was estimated that the biggest cartels ($C_1$ and $C_2$) suffered each fewer than 1,000 incapacitations in a year \cite{PrietoCampedelliHope2023}. Thus, most incapacitations do not directly reduce the size of the biggest cartels. The national impunity rate in Mexico stands at 96.3\%, with intentional homicide impunity reaching 95.7\% in 2022. In some cities, it rises to an alarming 100\% \cite{MexEvaluaImpunidad}. Security programs in Mexico could become more efficient at incarcerating criminals by prioritising the use of advanced police intelligence and focusing efforts on apprehending homicide perpetrators. By enhancing investigative capabilities and targeting those involved in severe crimes, law enforcement can shift resources from minor infractions to dismantling violent networks and addressing the root causes of insecurity. Better-trained and equipped police forces would lead to higher incapacitation rates for serious offences, ultimately strengthening public safety.
}

{
To analyse how sensitive the results are to the efficiency of security programs, we examine three budget scenarios (the current one, the zero budget and the optimal budget) and vary the efficiency of security programs with $\pi \pm 20\%$. Results show that with more efficient security programs, the number of homicides and cartel harm would drop faster with an unconstrained budget (Figure \ref{SensitivitySecurityPrograms}). However, results also show that in the long run, it is convenient to invest more resources in social and security programs, regardless of whether the efficiency of security programs varies $\pm 20\%$. 

\begin{figure}[!htbp]
\centering
\includegraphics[width=0.85\textwidth]{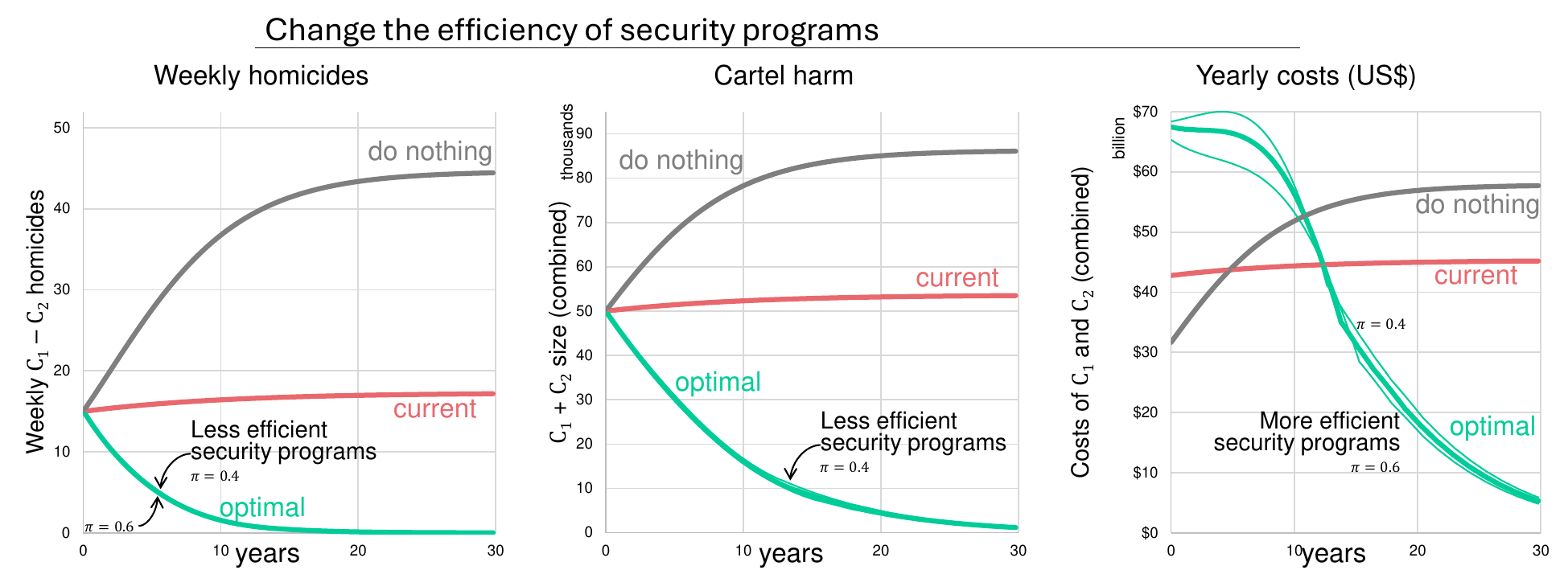}
\caption{Number of homicides, cartel influence and costs depending on the budget invested in social and security programs (by colour). The thin lines correspond to different values of the security program's efficiency.  }\label{SensitivitySecurityPrograms}
\end{figure}
}

{
Although we have simplified the security apparatus as a unified entity, Mexico operates within a federal system where state police forces interact with various federal agencies. Many of them no longer exist, such as the Federal Police and the National Gendarmerie, now replaced by the National Guard. These forces also interact with the judiciary and the penitentiary system. In recent years, several of these institutions have undergone a militarisation process, which has significant implications for how the country addresses cartel-related issues \cite{padilla2024militarization}. The political environment is also complex, with many states influenced by electoral cycles and often prioritising short-term outcomes over long-term strategies. This creates a tension between efficiency and optimisation, as governments frequently pursue multiple, often short-term, objectives \cite{kydland1977rules}.
}

\subsection*{M - Costs of the budget strategies}

The estimated 30-year costs of different budget allocation strategies are presented in Table \ref{costsPerStrategy}.

\begin{table}
    \centering
    \begin{tabular}{r | ccc | cc | cccc}
         & $u_1$ & $u_2$ & $v$ & 30-year cost  & \multicolumn{4}{c}{years for a 50 and 90\% drop} \\
Strategy & \text{US\$ } million & \text{US\$ } million & \text{US\$ } & \text{US\$ } billion &  \multicolumn{2}{c}{homicides} & \multicolumn{2}{c}{harm} \\
\hline
current & 21.43 & 21.43 & 15.94 & 804.48 &  NA & NA & NA & NA \\ 
\hline
do nothing & 0 & 0 & 0 & 933.26 & NA & NA & NA & NA \\ 
only security & 21.43 & 21.43 & 0 & 857.58  & NA & NA & NA & NA \\ 
only social & 0 & 0 & 15.94 & 878.62 &  NA & NA & NA & NA \\ 
\hline
optimal within budget& $u_1(t)$ & $u_2(t)$ & $v(t)$  & 798.39  & NA & NA & NA & NA \\ 
\hline
unconstrained budget& $u_1(t)$ & $u_2(t)$ & $v(t)$  & 719.0 & 3 & 8.4 & 5.6 & 14.8 \\ 
         \hline
    \end{tabular}
    \caption{Results of the budget strategies. NA means that the strategy does not reduce the homicides or cartel harm.}
    \label{costsPerStrategy}
\end{table}

\section*{Acknowledgements}

We sincerely thank Prof. Jonathan Caulkins for his insightful comments and valuable feedback, which have contributed significantly to the improvement of this work.

RPC is funded by the Austrian Federal Ministry for Climate Action, Environment, Energy, Mobility, Innovation and Technology (2021-0.664.668) and the Austrian Federal Ministry of the Interior (2022-0.392.231). 

\section*{Competing interests}

The authors declare that they have no competing interests.

\bibliographystyle{unsrt}

\end{document}